\documentclass[11pt,en]{elegantpaper}

\title{Integrating granular data into a multilayer network: an interbank model of the euro area for systemic risk assessment}

\author{
Ilias Aarab\thanks{Corresponding author: \email{Ilias.Aarab@ecb.europa.eu}.} \and
Thomas Gottron  \and
Andrea Colombo  \and
J\"org Reddig  \and
Annalaura Ianiro 
}

\institute{
European Central Bank\footnote{The views expressed in this paper are those of the authors and do not necessarily reflect those of the European Central Bank.}
}

\date{24 September 2023}

\usepackage{graphicx}
\usepackage{subcaption}
\usepackage{placeins}
\usepackage{booktabs}
\usepackage{amsmath,amssymb,amsthm}
\usepackage{setspace}
\usepackage{eurosym}

\usepackage{makecell}

\newcommand{\jel}[1]{\par\noindent\textbf{JEL:} #1}
\newcommand{\msc}[1]{\par\noindent\textbf{MSC:} #1}

\begin{document}

\maketitle

\begin{abstract}
Micro-structural models of contagion and systemic risk emphasize that shock propagation is inherently multi-channel, spanning counterparty exposures, short-term funding and roll-over risk, securities cross-holdings, and common-asset (fire-sale) spillovers. Empirical implementations, however, often rely on stylized or simulated networks, or focus on a single exposure dimension, reflecting the practical difficulty of reconciling heterogeneous granular collections into a coherent representation with consistent identifiers and consolidation rules. We close part of this gap by constructing an empirically grounded multilayer network for euro area significant banking groups that integrates several supervisory and statistical datasets into layer-consistent exposure matrices defined on a common node set. Each layer corresponds to a distinct transmission channel, long- and short-term credit, securities cross-holdings, short-term secured funding, and overlapping external portfolios, and nodes are enriched with balance-sheet information to support model calibration. We document pronounced cross-layer heterogeneity in connectivity and centrality, and show that an aggregated (flattened) representation can mask economically relevant structure and misidentify the institutions that are systemically important in specific markets. We then illustrate how the resulting network disciplines standard systemic-risk analytics by implementing a centrality-based propagation measure and a micro-structural agent-based framework on real exposures. The approach provides a data-grounded basis for layer-aware systemic-risk assessment and stress testing across multiple dimensions of the banking network.
\keywords{granular data collections, multi-layer network, financial contagion, systemic risk, micro-structural models}
\jel{D8, H51}
\msc{05C90, 68T30}
\end{abstract}

\section{Introduction}
\label{sec:intro}

A large literature studies how shocks propagate through financial networks and how amplification mechanisms can turn idiosyncratic stress into systemic events~\parencite{eisenberg2001systemic,boss2004network,montagna2016multi}. Micro-structural models emphasize that transmission is multi-channel: losses can spread through long-term counterparty exposures, roll-over risk in short-term funding markets, securities cross-holdings, and common-asset fire-sale spillovers~\parencite{allen2000financial,elliott2014financial,cifuentes2005liquidity}. Yet empirical implementations frequently operate on a single exposure dimension or on stylized network structures, largely because real-world bilateral exposures are fragmented across data collections, reported under different definitions and frequencies, and difficult to reconcile across entity identifiers and consolidation regimes.

In the last decade, policy makers and supervisors have gained access to an expanding set of granular statistical and supervisory collections that, in principle, enable a more realistic measurement of interconnectedness and risk transmission. Prominent examples include AnaCredit, EMIR, SFTDS, MMSR, SHS, FINREP and COREP\footnote{AnaCredit (analytical credit datasets) contains detailed information on individual bank loans above €25,000 to legal entities in the euro area, harmonised across all Member States. EMIR data underlies derivatives contract data collected under the European Market Infrastructure Regulation (EMIR). SFTDS is based on the EU SFTR (Regulation (EU) 2015/2365) to collect securities financing transaction data. The money market statistical reporting (MMSR) dataset provides information on the (un)secured, foreign exchange swap and overnight index swap euro money market segments. FINREP and COREP are European projects of the Committee of European Banking Supervisors (CEBS): FINREP stands for Financial Reporting and is based on the International Financial Reporting Standards (IFRS); COREP stands for Common Reporting of the solvency ratio and is based on the Capital Requirements Directive (CRD).}. A key challenge is to integrate these sources in a semantically consistent manner, including the alignment of identifiers, consolidation boundaries, instrument definitions, and reporting structures, and to map them into representations suitable for core analytical questions in macro- and micro-prudential surveillance.

A natural representation for such questions is a network (graph) in which nodes represent institutions (or groups under a given consolidation regime) and weighted directed edges represent exposures or claims. For many mechanisms, however, a single network is insufficient: different types of interactions (e.g., short- versus long-term credit, secured funding, or cross-holdings) encode distinct transmission channels and can exhibit markedly different topology~\parencite{bargigli2015multiplex,battiston2014structural}. Aggregating these channels into a single layer can therefore mask economically relevant structure and potentially misidentify the institutions that are systemically important in a particular market. Conversely, populating a multilayer model with real exposures is demanding: it requires both substantive knowledge of the reporting frameworks to resolve inconsistencies and scalable tooling to process large volumes of granular records.

This paper contributes to closing the gap between the multi-channel contagion mechanisms emphasized in the theoretical literature and what can be evaluated empirically on real systems. We construct an empirically grounded multilayer network for euro area significant banking groups by integrating several granular supervisory and statistical collections into layer-consistent exposure matrices defined on a common node set. The resulting network supports layer-aware stress-testing and provides a realistic empirical basis for systemic-risk analytics that are often studied on simulated networks. We then illustrate the implications of this measurement approach by implementing leading propagation-based and micro-structural agent-based frameworks on the constructed layers.

In particular we make the following contributions:
\begin{itemize}
	\item
	\textbf{Measurement:} We construct a multilayer exposure network for euro area significant banking groups by integrating multiple granular supervisory and statistical collections into a harmonized, consolidation-consistent representation with a common node set. We document the mapping, preprocessing, and aggregation steps that reconcile heterogeneous reporting schemes and identifiers.
	\item
	\textbf{Structure:} We provide evidence of pronounced cross-layer heterogeneity in connectivity and centrality and show that a flattened (aggregated) representation can conceal channel-specific structure that is relevant for propagation and systemic importance.
	\item
	\textbf{Implications for systemic-risk analytics:} Using real exposures, we implement two widely used frameworks, a centrality-based propagation measure and a micro-structural agent-based model, to illustrate how layer-aware measurement changes the identification and magnitude of systemic impact across channels, and to compare empirical outcomes to theoretical mechanisms.
\end{itemize}

The rest of the paper is structured as follows: We review related work in Section~\ref{sec:related}.
In Section~\ref{sec:data} we describe the data used and our data preparation and integration techniques.
Section~\ref{sec:modelling} details the construction of the multi-layer network, and
Section~\ref{sec:topology} provides compact topological diagnostics, with a more comprehensive treatment provided in \textcite{aarab2025topology}.
The analytical methods we apply on the network are presented in Section~\ref{sec:analysis}.
We conclude in Section~\ref{sec:summary}.

\section{Related work}
\label{sec:related}

The banking-network perspective has become a central empirical and theoretical tool for studying systemic risk, particularly in the wake of the 2008 financial crisis, which highlighted the difficulty of understanding and predicting propagation mechanisms in an interwoven financial system~\parencite{huser2015too}. A parallel development has been the gradual expansion of granular supervisory and statistical reporting, enabling more detailed measurement of interdependencies but also raising non-trivial integration challenges across heterogeneous data sources.

Related literature can be divided into three categories: (1) motivating work, arguing for the advantages of modelling financial markets as networks, (2) theoretical work, developing network models and dynamics to represent market behaviour and (3) empirical work, using different types of data to describe certain aspects of the network structures in the banking sector.

\cite{allen2009networks} argue that network theory can help in understanding financial systems, interactions among agents, and associated economic phenomena.
\cite{bandt2012systemic} consider the complex network of exposures among banks as one of the key features for determining systemic risk while \cite{di2018survey} review network-based approaches as an analytical basis for systemic risk indicators.
Recent work focuses in particular on multi-layer networks~\parencite{bargigli2015multiplex}. 
Multi-layer networks cater for different types of interrelation among banks and model them separately in distinct layers of the network, where each layer may exhibit a different topology~\parencite{battiston2014structural}.

Theoretical foundations of financial market network analytics address contagion chains caused by liquidity preference shocks~\parencite{allen2000financial}, cross-holdings~\parencite{elliott2014financial}, default cascades based on capitalisation and exposures in the network~\parencite{nier2007network} or contagion effects based on shocks to asset values~\parencite{glasserman2015likely}. 
Insights vary across models and assumptions.
A frequent conclusion is that highly connected nodes are often most exposed to spillover effects~\parencite{glasserman2015likely}.
\cite{nier2007network} observe that the effect of connectivity on contagion can be non-linear: initial increases in connectivity foster contagion, but beyond a threshold additional connectivity may dampen cascades by improving risk sharing. \cite{battiston2012liaisons} report similar non-linear behavior in networks of risk diversification.
\cite{elliott2014financial} model cross-holdings and investigate how defaults can trigger further failures.
\cite{glasserman2015likely} study contagion effects due to shocks to asset values, linking interconnectedness to expectations of losses and defaults.
\cite{cifuentes2005liquidity} provide a model of fire sales using a network based on overlapping portfolios: distressed entities liquidate assets, depressing prices and transmitting valuation losses to other holders.
DebtRank~\parencite{battiston2012debtrank} estimates the systemic importance of institutions based on balance-sheet buffers and network exposures, introducing domain-specific modifications of feedback centrality metrics to quantify distress propagation following exogenous shocks. The DebtRank value of a node corresponds to the additional distress in the system that is attributable to that node's initial default or distress.

Empirical work on interbank networks is largely data-driven, ranging from descriptive characterizations of topology to applications of contagion models.
\cite{boss2004network} analyze Austrian banks and document fat-tailed degree distributions, low clustering, and short average path lengths.
A recurring finding is a core--periphery structure, which often fits interbank data better than alternative generative mechanisms~\parencite{van2014finding} such as preferential attachment~\parencite{barabasi1999emergence}.
Similar observations are reported for German banks using bilateral exposure data~\parencite{craig2014interbank}, with interpretations linked to specialization and non-random formation of relationships.
An analysis of large EU banks~\parencite{alves2013structure} distinguishes between direct and indirect interconnections and studies temporal dynamics.

Empirical investigations of multi-layer networks capture different types of connections between banks. \cite{bargigli2015multiplex} distinguish contracts by maturity and secured/unsecured nature in the Italian market and show that layers can have sharply different topology; moreover, a flattened representation may fail to capture key aspects of complexity.
\cite{montagna2016multi} develop a micro-structural systemic-risk model on a multi-layer network combining short- and long-term loans with common exposures to financial assets, partially leveraging real-world data and filling gaps with simulated estimates.
\cite{aldasoro2018multiplex} use anonymized exposures among 53 large European banks, broken down by maturity and instrument type, to characterize multi-layer network features using data collected by the European Systemic Risk Board, the European Banking Authority, and national supervisory authorities.

Our contribution is complementary to these strands: Much of the systemic-risk literature develops micro-structural contagion models on stylized balance sheets and simulated networks, or studies a single channel of interdependence in isolation. We contribute an empirically grounded alternative that integrates multiple granular collections into a semantically consistent, multi-layer interbank network capturing distinct market segments and exposure types. On this integrated network, we extend widely used systemic-risk and contagion frameworks to operate directly on observed exposures and balance-sheet information, providing a unified basis to study how propagation patterns differ by market segment and how they interact across layers.

\section{Data}
\label{sec:data}

Granular datasets provide the foundation for creating our multi-layer networks. While Table~\ref{tab1} gives an overview of the various datasets we explain in the following their specificities and the rational for using them in our analysis.

\begin{table}[t]
\centering
\caption{Overview of granular datasets}
\label{tab1}
\begin{tabular}{@{}lccc@{}}
\toprule
Dataset & Available from & Frequency & Coverage / instrument \\ 
\midrule
AnaCredit & 2018--09 & Monthly   & Loan-by-loan information \\
CSDB      & 2009--04 & Monthly   & Security-by-security information \\
SHSG      & 2018--09 & Monthly   & Security-by-security holdings \\
SFTDS     & 2020--06 & Daily     & Repurchase and buy-sell back transactions \\
RIAD      & 1998--01 & Daily     & Banking group structures and identifiers \\
ROSSI     & 2019--01 & Monthly   & List of significant institutions \\
COREP     & 2017--06 & Semiannual& Capital information \\
FINREP    & 2016--09 & Quarterly & Financial information \\
\bottomrule
\end{tabular}
\end{table}

\textbf{Register of Institutions and Affiliates Data (RIAD):} RIAD is the shared master dataset serving several European System of Central Banks (ESCB) and Single Supervisory Mechanism (SSM) business processes and statistical data collections. The RIAD data model includes more than 100 properties for legal entities, including a wide range of entity identifiers and relationships (e.g. foreign branches and subsidiaries).
In our application, RIAD serves to identify entities across different datasets and as a source for constructing banking groups.

\textbf{List of significant banking groups (ROSSI list):}
Almost all available granular datasets comprise information on the largest banking groups in the EA and participating member states which are directly supervised by the SSM.
Hence, we use ROSSI to determine the sample of banking groups to be included in our multi-layer network.
As of 1\textsuperscript{st} of June 2021, this list of significant institutions comprised a total of 114 banking groups.
From ROSSI, we extract the RIAD codes of significant institutions, considering the prudential consolidation regime. This regime widens the group structure view as it also considers cases where the consolidating entity might be different than a bank (e.g. a financial holding) and improves consistency in the aggregation of granular datasets.

\textbf{CSDB and SHSG data:} The Centralised Securities Database (CSDB) 
is a reference database for all securities relevant for statistical purposes of the ESCB. It comprises information on debt securities, equity instruments and investment fund shares stored on a security-by-security basis. Each instrument is identifiable by its International Securities Identification Number. For each security, a large set of attributes is available.
While CSDB covers the issuance of securities by banking groups, the Securities Holdings Statistics Database by Group (SHSG) dataset contains the holdings of securities by banks. Thus, we also observe which banking groups hold issued securities, inside and outside the euro area.

\textbf{AnaCredit data:} The AnaCredit dataset~\cite{israel2017analytical}
contains loan-by-loan information collected from euro area banks extended to corporations. These data are reported at monthly frequency and are available as of September 2018. Among others, information on the outstanding amount, maturity, interest rate, collateral/guarantee, and on involved counterparties are collected for each individual loan. This makes AnaCredit a rich dataset for a wide range of analytical purposes.

\textbf{SFT – Securities Financing Transactions data}: The Securities Financing Transactions Data Store (SFTDS) collects and processes data reported under the Securities Financing Transactions Regulation (EU) 2015/2365\footnote{Regulation (EU) No. 2015/2365 on transparency of securities financing transactions and of reuse and amending Regulation (EU) No. 648/2012 Securities Financing Transactions Regulation – SFTR).}.
SFTs and the scope of the SFTR include\footnote{Article 3 (11) SFTR}:
(a) repurchase/ reverse repurchase transactions, (b) buy-sell back or sell-buy back transactions, (c) securities or commodities lending and securities or commodities borrowing and (d) margin lending transactions.

\textbf{FINREP/COREP:} FINREP (Financial Reporting Standards) and COREP (Common Reporting Standards) are the two reporting frameworks developed as part of the implementation of Basel III in Europe\footnote{EBA’s Implementing Technical Standards (ITS) amending the European Commission’s Implementing Regulation (EU) No 680/2014 on supervisory reporting of institutions under Regulation (EU) No 575/2013.} which gives the European Banking Authority a mandate to request both capital and financial information.
COREP specifies the framework related to capital information, while FINREP specifies the financial information.
Banks subject to International Financial Reporting Standards (IFRS) already use FINREP templates to submit financial information in a harmonised format at consolidated level. This requirement has also been extended to SSM banks reporting at sub-consolidated or solo level under IFRS and national Generally Accepted Accounting Principles (GAAPs)\footnote{Regulation (EU) 2015/534 of the ECB of 17 March 2015}. FINREP templates are subject to quarterly mandatory reporting.
COREP is used to collect Pillar 1 data and information on liquidity, leverage and large exposures from banks in a harmonised format.

\section{Network modelling}
\label{sec:modelling}

\begin{figure}[tb]%
\centering
\includegraphics[width=0.85\textwidth]{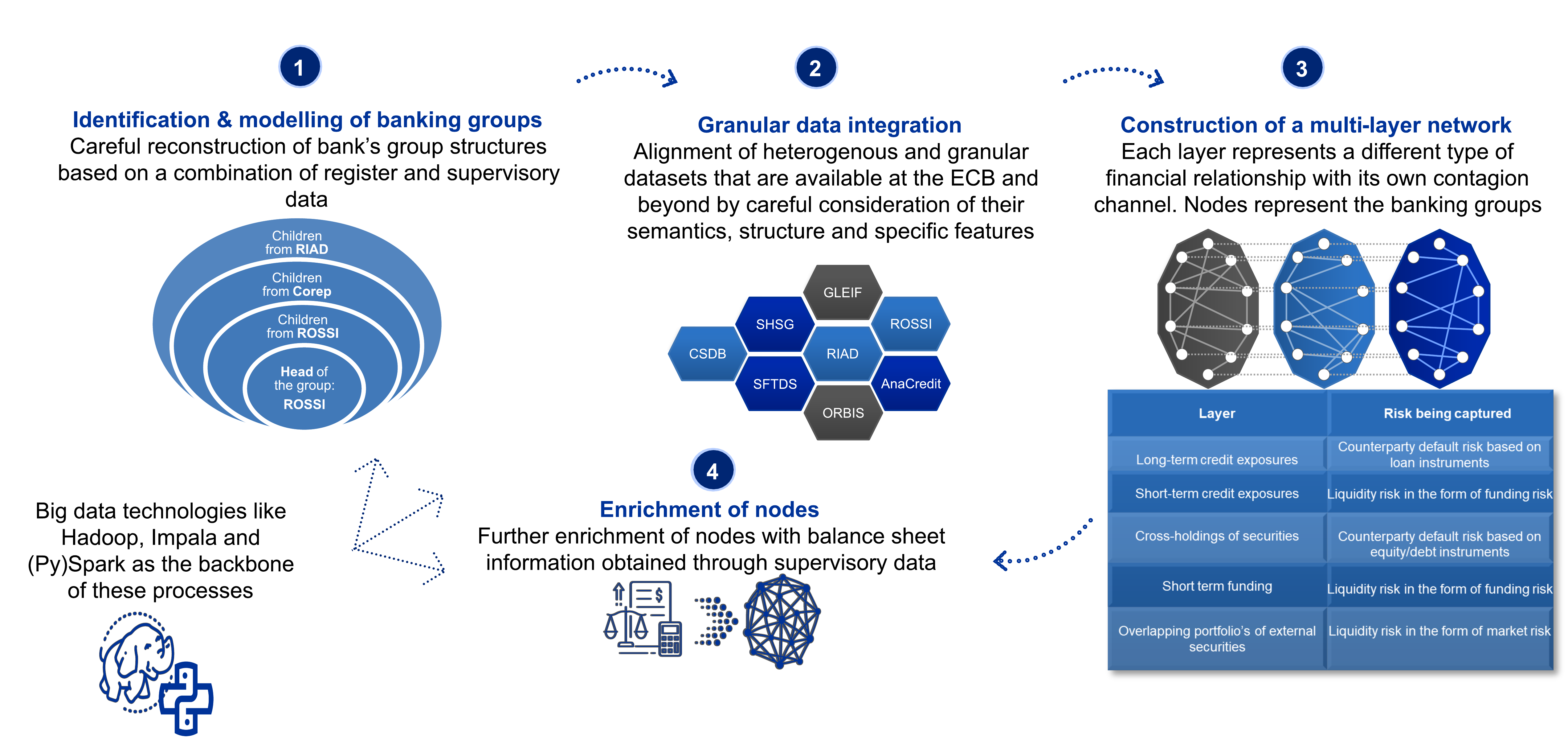}
\caption{1. Using ROSSI, RIAD and COREP data a list of banking groups, including their group structure is constructed. 2. A multitude of granular datasets are integrated with banking groups. 3. A multi-layer network is constructed, with nodes depicting banking groups, layers depicting financial markets and edges representing financial relationships between nodes. 4. Further enrichment of the nodes is carried out to represent nodes by their balance sheet.}\label{modelling_overview}
\end{figure}

This section provides details on how we construct our multi-layer network based on the various datasets. 
Figure~\ref{modelling_overview} provides a bird eye view of the modelling process.\footnote{Note that the analysis we conduct focuses on a static snapshot of the multi-layer network. The considered snapshot is the end-of-month observation of the network as of June 2021. This is a date in which all datasets, considering their respective frequencies, are available. }

\subsection{Identification and modelling of banking groups}\label{mod1}

The first step in modelling a multi-layer network is to define the nodes $n_i$ of the network. As our focus is to measure exposures between banking groups across different markets, we model banking groups as nodes. The construction of banking groups is initiated from the head of a banking group. The list of group heads of significant\footnote{The definition of Significant Institutions can be found in Regulation (EU) No 468/2014 of the European Central Bank of 16 April 2014.} banks in the EA is retrieved from ROSSI.
Once the group head has been identified, all entities belonging to that group need to be identified.
We consider all subsidiaries as being part of the group over which the group head exercises direct or indirect control, based on equity share\footnote{In addition, supervised entities according to Guideline (EU) 2020/497 of the ECB are also included in the group, if not already included following the control relationship criteria.}.
To obtain a complete group structure, including subsidiaries that are not credit or financial institutions, we leverage and integrate different group information available in ROSSI, COREP and RIAD.

\subsection{Granular data integration}\label{mod2}

After the identification of banking groups and their subsidiaries, we integrate the different granular datasets to form layer-specific exposures. Consistent identification of group entities is realized using entity identifiers available in RIAD. During the integration process we harmonize and align different structures and semantics of the datasets (e.g., stock-based versus transactional reporting).
Having information on the banking group structure allows us to aggregate subsidiary-level records to the group-head level and to identify interrelations across all granular datasets\footnote{To efficiently process the large amount of data we make use of a Hadoop based data lake to store the data and utilize Spark for distributed processing~\cite{aarab2022analysis}.}.

\subsection{Construction of a multi-layer network}\label{mod3}

We follow~\cite{battiston2014structural} and define a multi-layer network as \emph{a network where each node appears in a set of different layers $l$, and each layer describes all the edges of a given type}. This entails that the layers share the same node set. In our case, edges $e_{ij}^l$ from node $n_i$ to node $n_j$ are directed and their weight indicates the aggregated exposure of banking group $i$ towards $j$ within layer $l$\footnote{Note that intra-group exposures of banking groups are removed to avoid self-loops within the network.}. The edge values differ across layers $l$ based on the type of exposure they represent:

\textbf{Long-term credit layer}: This layer contains loan exposures between banking groups having an initial maturity of at least three months, in line with~\cite{montagna2016multi}. Exposures are aggregated over different types of loan instruments (e.g., deposits, credit lines, convenience credit, etc.), with the bulk (more than 90\%) of exposures coming from deposits, credit lines other than revolving credit and unspecified loans. Edge weights $w_{ij}^{\textit{ltc}}$ indicate the outstanding nominal values.

\textbf{Short-term credit layer}: This layer contains loan exposures with an initial maturity shorter than three months. Exposures are processed in the same way as for the long-term credit layer, with the additional constraint of ensuring strictly positive maturities of the loans. We observe that a large part of instruments (more than 45\%) are reverse repurchase agreements, confirming the short-term nature of the layer. Edge weights $w_{ij}^{\textit{stc}}$ indicate the outstanding nominal values.

\textbf{Cross-securities layer}: This layer contains equity and debt holdings between banking groups. An edge from node $n_i$ to node $n_j$ represents the market value of the securities issued by $n_j$ and held by $n_i$. We do not distinguish between equity and debt securities and aggregate them into the same edge. The edge weight $w_{ij}^{\textit{cs}}$ represents the market value of investments made by $n_i$ into $n_j$’s issued securities.

\textbf{Short-term funding layer}: The short-term funding layer contains repurchase agreements and buy-sell back transactions obtained from the SFT dataset. Within this layer the edge weight $w_{ij}^{\textit{stf}}$ from node $n_i$ to $n_j$ represents the aggregated open funding transactions between banking group $n_i$ and $n_j$. The edge $e_{ij}^{\textit{stf}}$ is directed from the collateral taker $n_i$ to the collateral giver $n_j$. To align with the end-of-month snapshot used across layers, we extract transactions that are indicated as active at the end of the month.


\textbf{External securities layer:}
The external securities layer models investments by banking groups in securities issued by entities outside the interbank market. A natural representation is a bipartite network with two disjoint node sets: banking groups and external securities. Directed edges from a banking group to a security encode the market value of the corresponding holding. In our application, however, the primary interest is in the extent to which this layer can generate fire-sale spillovers. Specifically, when a banking group $i$ liquidates a substantial fraction of its securities portfolio, we ask how this may affect the balance sheets of other banking groups through common-asset holdings. To capture this mechanism parsimoniously while preserving a common vertex set across layers, we use the projected representation on the banking-group node set \(N\): we construct an undirected graph in which an edge between banking groups $i$ and $j$ represents the market value of overlap in their portfolios.

For example, suppose banking group $i$ holds $5$ shares of security $s_{1}$ and banking group $j$ holds $3$ shares of $s_{1}$, where the market price of $s_{1}$ is \EUR{100}, and the two groups hold no other common securities. Then the edge weight between $i$ and $j$ equals the market value of the overlapping position, i.e.,
\[
w_{ij} \;=\; \min(5,3)\times 100 \;=\; 300 \, .
\]
Under this construction, the impact of a sell-off by $i$ can be traced by following edges that are directly or indirectly connected to $i$: banking groups linked to $i$ are more likely to experience balance-sheet deterioration when $i$ liquidates large positions in overlapping securities. To limit the universe of securities, we aggregate holdings at the issuer level, which also has the advantage of implicitly accounting for positive correlations among securities issued by the same issuer. Data are derived mainly from SHSG and mapped to banking groups through RIAD and ROSSI.

\textbf{Flattened network layer:}
The flattened network layer is an artificially constructed layer, in the sense that it does not correspond to a specific real-world market. Instead, it is defined as the sum of all layers in the multi-layer network, where summation corresponds to edge-wise aggregation. If an edge from banking group $i$ to $j$ appears in multiple layers, its weights are added; if an edge appears in one layer but not another, it is carried over unchanged. The same logic extends to the aggregation of more than two layers. Care is required when adding undirected layers (in our case, the external securities layer). To obtain a meaningful aggregated \emph{directed} layer, we first convert each undirected edge into a pair of symmetric directed edges (i.e., we construct a directed graph in which, for every edge, the reverse edge is also present) before summing it with the other layers. This ensures that the network effects induced by this layer are represented in the flattened network (i.e., price deterioration can propagate in either direction between two connected banking groups). At the same time, edge weights in the flattened network lose a direct economic interpretation because aggregation can introduce double counting (e.g., the total weight in the flattened network double counts the market value of overlapping portfolios). The flattened network provides a useful benchmark to compare layer-specific structures with the system as a whole and to assess the contribution of each layer to aggregate network connectivity. By construction, it draws on all datasets described above.

\subsection{Enrichment of the nodes}\label{mod4}

After constructing the multi-layer network, we enrich nodes with key balance-sheet items (e.g., Tier 1 capital and total assets) to support calibration and interpretation of systemic-risk measures. Total assets are retrieved from FINREP\footnote{Total assets are available in FINREP template: F 01.01 - Balance Sheet Statement.} and Tier 1 capital from COREP\footnote{Tier 1 Capital is available in the COREP template: C47.00 – Leverage ratio calculation. The amount of Tier 1 capital is calculated according to article 25 of the CRR, without taking into account the derogation laid down in Chapters 1 and 2 of Title I of Part Ten of the CRR.}. These values are available at consolidated level, enabling a straightforward mapping to banking groups via RIAD identifiers. The enriched nodes allow a more economically interpretable analysis of propagation and systemic impact across layers.

\section{Topological diagnostics}\label{sec:topology}

A comprehensive topological characterization of the constructed multi-layer interbank network
(e.g., layer-specific global metrics, tail behavior, and cross-layer comparisons) is provided in
\textcite{aarab2025topology}. In this paper, we report two compact diagnostics that (i) validate that
the constructed layers exhibit the pronounced heterogeneity commonly documented for interbank networks,
and (ii) motivate conducting the financial applications in Section~\ref{sec:analysis} in a layer-aware
manner rather than relying solely on an aggregated representation.

\subsection{Degree heterogeneity across layers}\label{sec:topo_degree}

\begin{figure}[!htbp]
\centering
\includegraphics[width=0.95\textwidth]{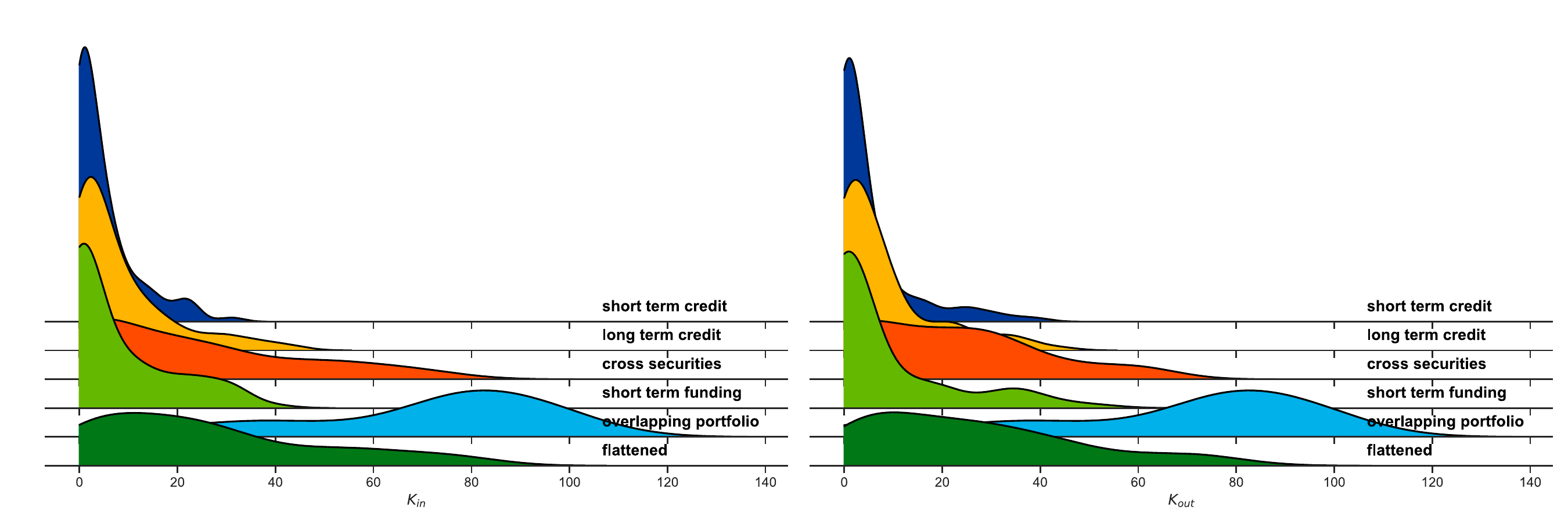}
\caption{\textbf{Empirical degree distributions across layers (June 2021 snapshot).}
Left panel: in-degree distributions; right panel: out-degree distributions (x-axes depict the number of edges).
Layers are color-coded: short-term credit (dark blue), long-term credit (yellow), cross-securities (red),
short-term funding / repo market (light green), external securities (light blue, dubbed overlapping portfolio), and the flattened (aggregated) layer (dark green).
Density estimates are obtained via a Gaussian kernel density estimator with bandwidth selected according to Scott's rule \parencite{scott1992multivariate}.}
\label{fig:deg_distributions}
\end{figure}

Figure~\ref{fig:deg_distributions} compares empirical in- and out-degree distributions across layers.
Two patterns are particularly relevant for the subsequent systemic-risk analytics.
First, within each layer, the in- and out-degree distributions are closely aligned, indicating a broadly balanced system
with both active creditors and debtors.
Second, the degree distributions differ markedly across layers, indicating that the same set of banking groups
is embedded in market segments with substantially different connectivity profiles.
In line with prior empirical evidence on interbank networks, the credit and funding layers display pronounced right-skewness
and fat tails that are visually compatible with scale-free-type heterogeneity~\parencite{boss2004network,craig2014interbank,van2014finding}, whereas the cross-securities layer appears
substantially flatter (closer to a uniform shape), and the external-securities layer exhibits a distinct left-skewed fat-tailed
structure (with a median centred around roughly 76 edges, consistent with its largest connected subcomponent).
Third, the aggregated (flattened) representation can mask this layer-specific heterogeneity: depending on how
layers are combined, the resulting degree shape may be dominated by the most populated layer.
In our data, the flattened layer aligns most closely with the degree distribution of the cross-securities layer,
despite being a composite of credit and funding markets as well.
This observation supports a key modelling choice of our application section: contagion and systemic impact are evaluated
on the relevant layers (or controlled combinations of layers), rather than relying on a single aggregate network as a proxy
for all market mechanisms.

\subsection{Centrality and core--periphery structure}\label{sec:topo_centrality}

\begin{figure}[!htbp]
\centering
\includegraphics[width=0.95\textwidth]{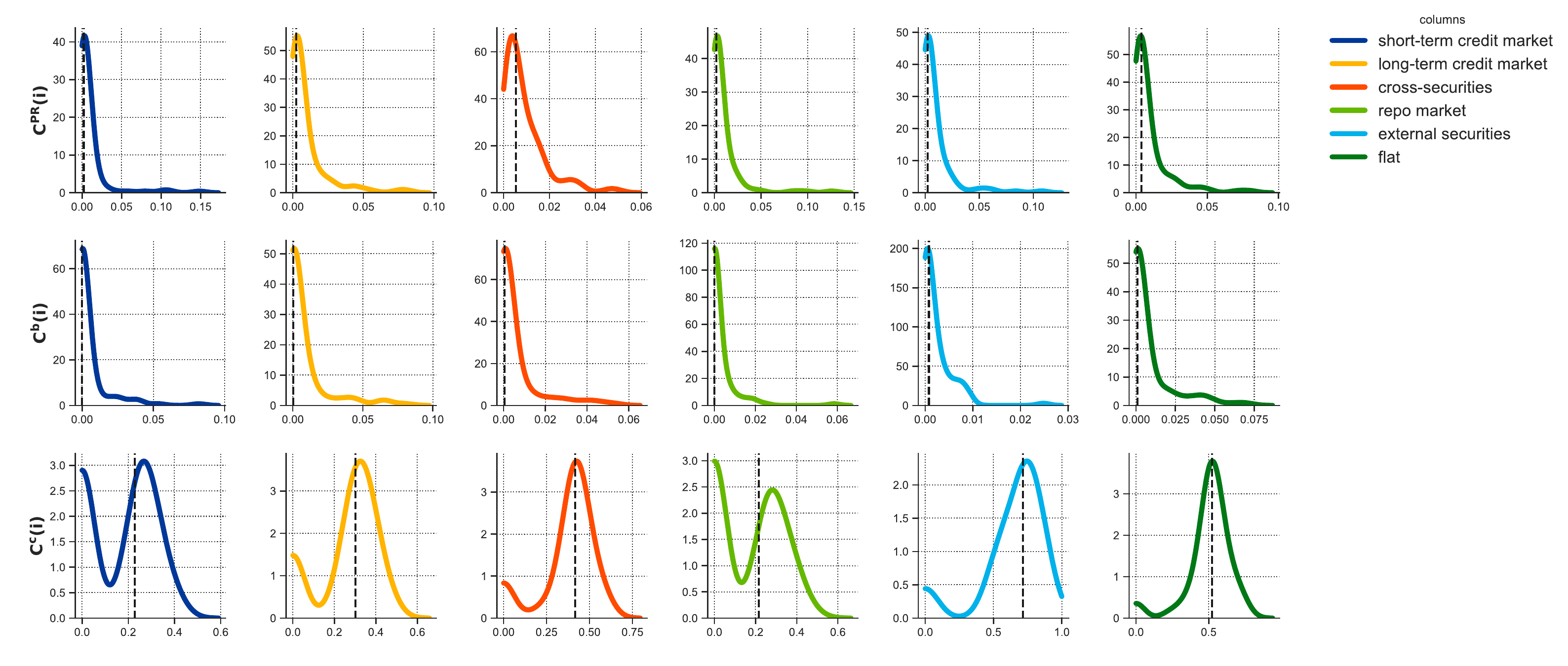}
\caption{\textbf{Centrality distributions across layers (June 2021 snapshot).}
Rows depict (normalized) centrality measures: PageRank \(C^{PR}(i)\), betweenness \(C^{b}(i)\), and closeness \(C^{c}(i)\).
Columns represent layers; x-axes depict centrality values normalized to \([0,1]\).
Layers are color-coded: short-term credit (dark blue), long-term credit (yellow), cross-securities (red),
short-term funding / repo market (light green), external securities (light blue), and the flattened (aggregated) layer (dark green).
The black vertical dashed line in each subplot denotes the median for the corresponding layer--measure pair.
Density estimates are obtained via a Gaussian kernel density estimator with bandwidth selected according to Scott's rule \parencite{scott1992multivariate}.}
\label{fig:centrality_distributions}
\end{figure}

Figure~\ref{fig:centrality_distributions} reports distributions of standard centrality measures across layers.
While most institutions concentrate at low centrality values, the distributions also indicate that a comparatively small subset
of nodes occupies structurally important positions (e.g., short-path accessibility and intermediation roles).
Across layers, PageRank~\parencite{page1999pagerank} and betweenness centralities concentrate near small values (with most mass between 0--5\% and maxima barely exceeding
the 10\% mark), suggesting that only a limited subset of institutions can strongly intermediate or steer propagation paths.
In contrast, closeness centrality exhibits a bimodal structure for most layers, with one mode near zero and a second mode typically between
25--75\% depending on the layer.
This pattern is consistent with a core--periphery organization: a relatively small and well-connected core coexists with a large periphery that is weakly embedded.
For the financial applications in Section~\ref{sec:analysis}, this provides a structural rationale for two recurring themes:
(i) systemic impact is driven not only by balance-sheet size but also by network position, and (ii) propagation dynamics can be
sensitive to shocks affecting core institutions even when average connectivity appears moderate.\footnote{In- and out-degree centralities are not shown, as their distributions closely mirror the in- and out-degree patterns in Figure~\ref{fig:deg_distributions}.}

\section{Layer-aware systemic-risk assessment}
\label{sec:analysis}

We use the constructed multilayer network to address financial-stability questions for euro area banks. Our aim is not to propose a new contagion channel, but to move beyond single-layer or stylised calibrations by extending leading network-based frameworks to a setting in which exposures are empirically observed across multiple market dimensions and can be analysed in a layer-consistent way. This provides an empirical basis to evaluate how systemic impact and propagation patterns depend on the market segment through which stress travels, and how conclusions can change when the same institutions are embedded in different layers. Building on the literature’s shock-based analytics for systemic risk, we focus on two widely used but complementary approaches: a propagation-based systemic-importance measure and a micro-structural agent-based framework. 

\subsection{Notational preliminaries}\label{mod5}

Throughout this section we utilize a number of definitions and notational
conventions derived from Linear Algebra and Graph Theory. The
conventions are standard and utilized commonly in financial literature
as well. However, as we deal with multiple graphs, multiple variables,
and different dimensions, we collect these definitions together in this
section to make referencing throughout the paper easy.

Let \(\Omega : = (N,\ G)\) denote a multi-layer network where \(N\) is a
set of nodes with cardinality \( \lvert N \rvert : = n = 114\), equal
to the number of Significant Institutions in our network. \(G\) denotes
a set of graphs making up the different layers of the network. The
cardinality of \(\lvert G \rvert : = L\), represents the number of
layers in the network. A graph, \(g_{l}\), related to a layer
\(l \in G\) can typically be represented by the tuple \((N,\ W_{l})\),
with \(W_{l} \in \mathbb{R}^{N \times N}\) an adjacency matrix
representing the directed and weighted edges in the layer. Elements of
\(W_{l}\), \(w_{\text{ij}}^{l}\) with \(i,j \in 1,\ \ldots,\ n\),
represent the weight of the edge coming from node \(n_{i}\) towards node
\(n_j\). \(w_{\text{ij}}^{l}\)= 0 signifies that no edge is present that
goes from \(n_{i}\) to \(n_j\) within layer \(l\). If not explicitly mentioned
otherwise, the direction of an edge in a graph follows the flow
of funds. That is \(w_{\text{ij}}^{l}\) represents a notion of the
exposure of \(n_{i}\) towards \(n_j\), with the exact definition depending on
the layer we are considering.

For the external securities channel, we distinguish between (i) the security-level representation used to model fire-sale price dynamics and (ii) the projected network-layer representation used in the multi-layer graph \(\Omega\). In the security-level representation, we work with a securities universe \(M\) of cardinality \(m\) and a holding matrix \(S \in \mathbb{R}^{n \times m}\), where \(s_{i\mu}\) denotes the amount of security \(\mu \in M\) held by banking group \(n_i \in N\). In the multi-layer network \(\Omega\), the external securities layer itself is represented on the common node set \(N\) via the projected overlap graph (an undirected layer), where edge weights capture the market value of overlap in portfolio holdings. Note that for all graphs in \(G\) we have
that the set of nodes is defined by the same set \(N\), underlying one
of the key features of a multi-layered network. The multi-layer network
is thus composed of a set of layers representing financial markets which
in turn are mathematically represented as graphs.

Every node in the network is characterised by an internal structure
given by its (consolidated) balance sheet. Balance sheet items of the
nodes are typically represented within an \(n\)-dimensional Euclidean
vector space \(\mathbb{R}^{n}\) with elements of the vector depicting
the balance sheet item value for the different nodes in the system
(e.g., for equity we have the vector \(Eq \in \mathbb{R}^{n}\) with
\(eq_{i}\) the equity value of node \(n_{i}\)). Similarly
state vectors representing the state of a node within the network at any
given time are represented as \(\left\lbrack a,\ b \right\rbrack^{n}\)
with \(a\) and \(b\) respectively the lower- and upper bound levels that
the vector contains. Binary state vectors are depicted as
\(\left\{ 0,1 \right\}^{n}\). A vector of ones is depicted as
\(\mathbf{1}_{\mathbf{n}} \in \mathbb{R}^{n}\). The transpose of a
matrix or vector \(V\) is given by \(V^{T}\), with the transpose of a
vector indicating a column vector.

The matrix multiplication is not defined by an explicit operator but can
be inferred by the juxtaposition position of two matrices/vectors
without any intermediary operation symbol, e.g., the matrix
multiplication between \(A\) and \(B\) is represented by \(\text{AB}\).
The elementwise Hadamard product is defined as \(A \cdot B\), with
broadcasting of vectors inferred from the dimensions of matrices being
multiplied with the vector. The Hadamard product between a matrix and a
row vector will be broadcasted row-wise, whereas for a column vector
broadcasting is done column-wise. The \(\min{\{\}}/max\{\}\) operators
are always defined element wise,
\begin{align}
\min\left\{ A \in \mathbb{R}^{n},\ B \in \mathbb{R}^{n} \right\} : = \left( \min\left\{ a_{1},\ b_{1} \right\},\ \ldots,\min\left\{ a_{n},\ b_{n} \right\} \right) \in \mathbb{R}^{n}\nonumber \\
\max\left\{ A \in \mathbb{R}^{n},\ B \in \mathbb{R}^{n} \right\} : = \left( \max\left\{ a_{1},\ b_{1} \right\},\ \ldots,\max\left\{ a_{n},\ b_{n} \right\} \right) \in \mathbb{R}^{n}\nonumber
\end{align}

\subsection{Propagation-based systemic importance}\label{6.1} 

We apply the original DebtRank~\parencite{battiston2012debtrank} algorithm to the multilayer network in two distinct calibrations: (a) a counterparty-loss channel, capturing credit risk, and (b) a liquidity-shortfall channel, capturing roll-over and funding risk. The objective is to quantify how a bank's distress can propagate through a given layer and to compare systemic impact across channels.

The credit-risk calibration is based on a vector $ Eq \in \mathbb{R}^{n}$ of banks' capital buffers to absorb shocks\footnote{To construct $Eq$ we make use of the Tier 1 capital information of banks, as proposed by~\cite{battiston2012debtrank}.}.
The exposures of banking group $n_i$ towards its interbank debtors relative to its capital buffer $eq_{i}$ is described by $w_{ij}^{\textit{eq},l_1} = \min\left( 1,\frac{w_{ij}^{l_1}}{\textit{eq}_{i}} \right)$ for $l_1 \in \{\textit{ltc}, \textit{cs}\}$, i.e., the long-term credit market and the cross-securities market.
The edge weight $w_{ij}^{\textit{eq},l_1}$ represents the potential loss that banking group $n_i$ can suffer when banking group $n_j$ defaults (assuming no recovery of debt is possible in the short run) as a fraction of its capital buffer. Note that if a potential loss from the exposures of banking group $n_i$ exceeds its capital buffer, then $w_{ij}^{\textit{eq},l_1} = 1$, representing the maximum loss that $n_i$ can experience.
We simulate distress propagation within the long-term credit and cross-securities layers by applying the DebtRank recursion.

Similarly, the liquidity-shortfall calibration uses a vector $Liq \in \mathbb{R}^{n}$ of banks' liquidity buffers to absorb shocks.
For the liquidity buffers we use the cash (asset) and deposits (liabilities) positions of a banking group $n_i$ to define its buffer $\textit{liq}_{i} = \textit{Cash}_{i} - \beta \cdot \textit{Deposits}_{i}$, where the parameter $\beta \in \lbrack 0,1\rbrack$ is a liquidity buffer scaler.
In our experiments we set $\beta$ to 0.05, 0.1 and 0.2 to examine different possible scenarios~\parencite{montagna2016multi, brandi2018epidemics}.
In this way we investigate the Liquidity Coverage Ratio of a banking group under different stressed circumstances, with higher $\beta$ values indicating a higher net cash outflow. The exposure is defined by $w_{ij}^{\textit{liq},l_2} = \min \left( 1, \frac{w_{ji}^{l_2}}{\textit{liq}_{i}} \right)$ for $l_2 \in \{\textit{stc},\textit{stf}\}$, i.e., the short-term credit and the short-term funding market.
$w_{ij}^{\textit{liq},l_2}$ represents the potential liquidity shortfall that banking group $n_i$ can suffer when banking group $n_j$ defaults as a fraction of its liquidity buffer. Again, if a potential liquidity shortfall exceeds the liquidity buffer, then $w_{ij}^{\textit{liq},l_2} = 1$, representing maximum depletion.

Next, we have a state vector
\(H \in \left\lbrack 0,1 \right\rbrack^{n}\) representing the distress
level of the banking groups, with a level of \(h_{i} = 1\) indicating the default
of banking group \(n_{i}\). We also denote with \(D \in \left\{ 0,1 \right\}^{n}\) a
binary state vector which represents whether a banking group \(n_{i}\) is in distress
(\(d_{i} = 1 \Leftrightarrow h_{i} > 0\)) or not
(\(d_{i} = 0 \Leftrightarrow h_{i} = 0\)). Lastly, we have the binary
state vector \(I \in \left\{ 0,1 \right\}^{n}\) indicating the set of
inactive nodes. Nodes in a distressed state become inactive nodes in the
next period, to ensure no infinite reverberations are introduced into
the system. With this in mind we can simulate the dynamics of a shock
propagation within the long-term credit market or cross-securities
market as follows:

\begin{align}
H\left( t \right)
&=
\min\left\{
H\left( t - 1 \right)
+ W^{l_1,\textit{eq}}
\left[\, H(t - 1) \cdot D(t - 1)\,\right],
\ \mathbf{1}_{\mathbf{n}}
\right\},
\\
I\left( t \right)
&=
\min\left\{
I\left( t - 1 \right) + D\left( t - 1 \right),
\ \mathbf{1}_{\mathbf{n}}
\right\},
\\
d_{i}\left( t \right)
&=
\begin{cases}
1, & \text{if } h_{i}\left( t \right) = 1 \text{ and } I_{i}\left( t - 1 \right) = 0,\\
0, & \text{otherwise.}
\end{cases}
\end{align}

For the
liquidity shortfall propagation, we have a very similar dynamic with the
exception that the matrices \(W^{l_1,\textit{eq}}\) are now replaced
by \(W^{l_2,\textit{liq}}\).

The above equations define a recursive propagation mechanism. Distress at time \(t-1\) is transmitted through the
weight matrix \(W^{l_1,\textit{eq}}\) (or \(W^{l_2,\textit{liq}}\) for the liquidity layers), updating the distress levels \(H(t)\).
The binary vector \(D(t)\) flags distressed nodes, while \(I(t)\) marks nodes that become inactive after being distressed,
preventing repeated reverberations. The process converges once no additional nodes enter distress.

To trigger the dynamics, we have the following set of initial
conditions:

\begin{itemize}
\item
  \(h_{i \in S_{f}\ }\left( t = 1 \right) = 1\) and
  \(h_{i \notin S_{f}\ }\left( t = 1 \right) = 0\)
\item
  \(d_{i \in S_{f}\ }\left( t = 1 \right) = 1\) and
  \(d_{i \notin S_{f}\ }\left( t = 1 \right) = 0\)
\end{itemize}

with \(S_{f}\) the singleton set containing the node that is going
initially in default due to an exogenous shock introduced to the system.

After the dynamics converges at time \(T\), we can compute the DebtRank
value of node \(n_i \in S_{f}\) as:

\[DR_{i} = \ \sum_{j}^{}{\left\lbrack h_{j}\left( t = T \right) - h_{j}\left( t = 1 \right) \right\rbrack v_{j}}\]

With
\(v_{i} = \frac{\sum_{j}^{}W_{\text{ij}}^{l}}{\sum_{i}^{}{\sum_{j}^{}{W_{i,j}^{l}\ }}}\)
a proxy for the relative economic value of a node defined as the
proportion of total investments (funding for the liquidity layers)
contributed by the node \(n_{i}\). \(DR_{i}\) thus measures the additional
distress propagated into the system, excluding the initial distress
caused by the exogenous shock. This additional distress can be used as a
proxy for the systemic impact of \(n_{i}\) on a specific layer within
the multi-layer network. For \(l_1 \in\) \{long-term credit
market, cross-securities market\}, systemic impact reflects potential default cascades due to counterparty risk.
For \(l_2 \in\) \{short-term credit market,
short-term funding market\}, systemic impact reflects potential liquidity crunches due to roll-over and funding risk.\footnote{The DebtRank modifications to capture liquidity shortfalls are partly based on the Exposed--Distressed--Bankrupted (EDB) model of~\cite{brandi2018epidemics}, with the main difference that we incorporate information on banks' balance-sheet composition, as suggested as an extension by the authors.}

For each node $n_i$ of the network, we initiate both models to compute its DebtRank. Comparing across layers directly addresses the multi-channel nature of contagion: institutions that appear marginal in one market can be central in another, and the ranking of systemic importance can shift with the transmission mechanism~\parencite{bargigli2015multiplex}.

\begin{figure}[!htbp]
\centering
\includegraphics[width=0.9\textwidth]{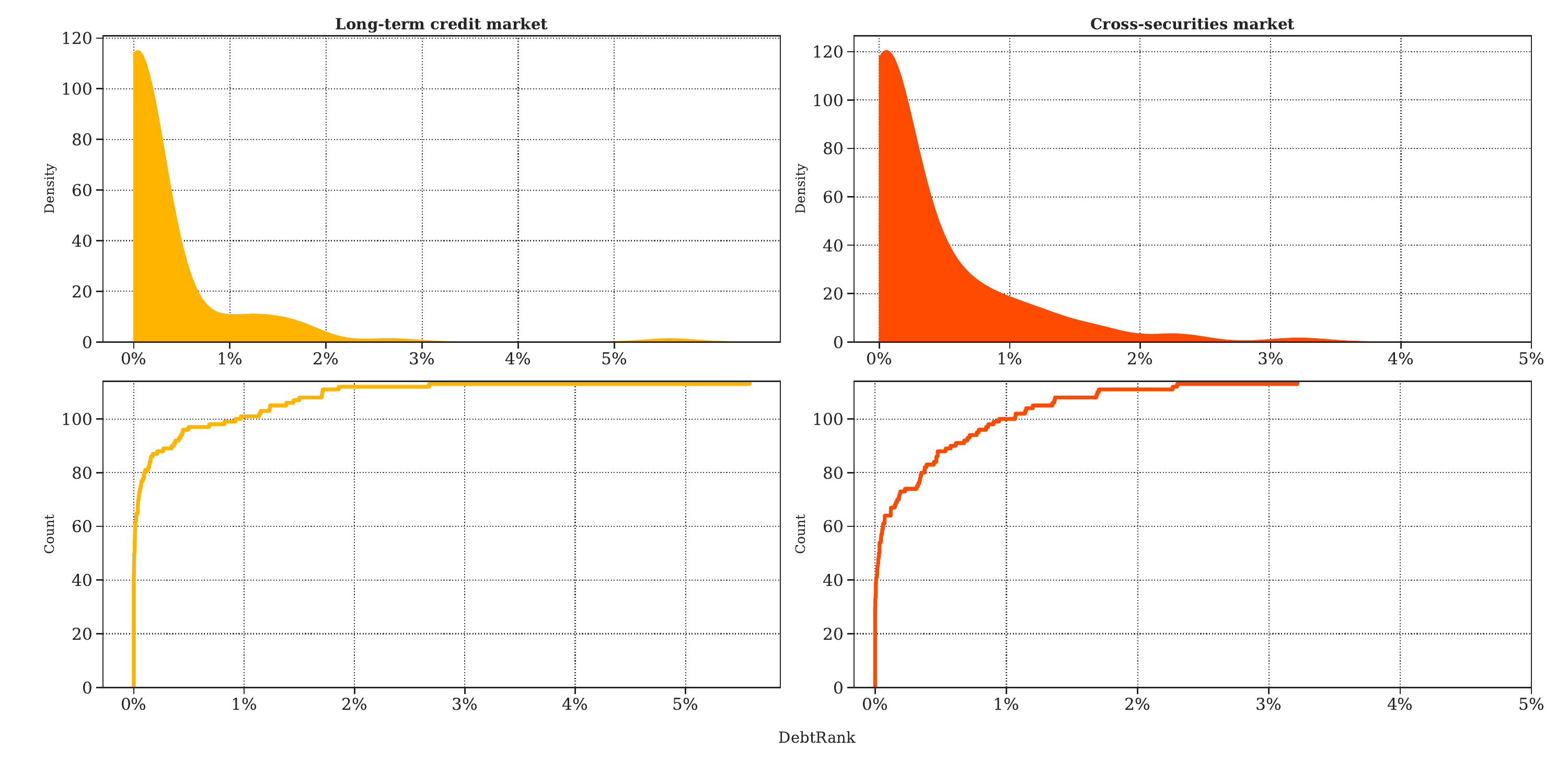}
\caption{Top panel: densities of DebtRank values for the long-term credit
layer (left in yellow) and cross-securities layer (right in red); x-axes
depict DebtRank values in (\%).
Bottom panel: cumulative counts of the same DebtRank values for the two layers}
\label{DR_default_risk}
\end{figure}

Figure~\ref{DR_default_risk} shows the results for the credit risk simulation.
For both the long-term credit market and the cross-securities market we observe relatively small DebtRank values with the bulk below 2\%. This implies that most banks would account for less than 2\% of the system's economic value in the event of their default. In the long-term credit layer, the most systemic bank accounts for roughly 5\% of the market, whereas in the cross-securities layer the maximum is around 3\%. Importantly, the identity of the most systemic bank differs across these two layers, consistent with the notion that systemic importance is channel-specific in multilayer financial networks~\parencite{bargigli2015multiplex}.

\begin{figure}[!htbp]
\centering
\includegraphics[width=0.9\textwidth]{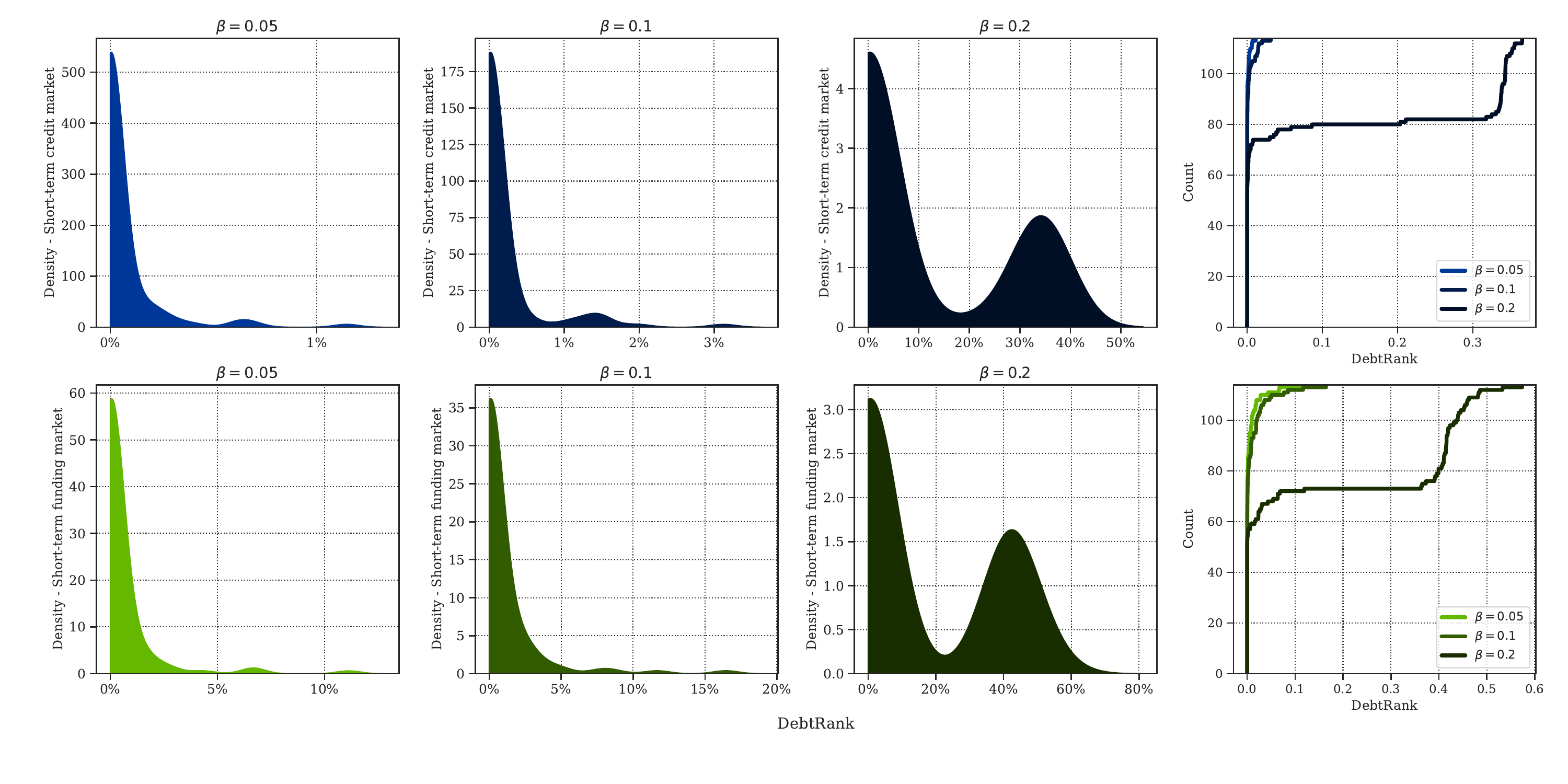}
\caption{
DebtRank values for the two layers that are concerned with
funding risk, the short-term credit market (top panel, color-coded in
blue shades) and the short-term funding market (bottom-panel,
color-coded in green shades). First three columns represent DebtRank
values for respectively \(\beta = 0.05\), \(\beta = 0.1\) and
\(\beta = 0.2\), with \( \beta \in [0,1] \) a liquidity buffer scaler with
higher \(\beta\) values aligned with a higher net cash outflow. The last column
shows the cumulative counts of DebtRank (darker shades indicate higher
\(\beta\) values). x-axes depict DebtRank values in (\%) .
}\label{DR_liq}
\end{figure}

Likewise, Figure~\ref{DR_liq} reports DebtRank values for the liquidity-shortfall calibration across different values of the liquidity buffer scaler \(\beta\).
A first pattern is the substantial difference between the two short-term layers: the repo market exhibits materially larger propagation than the short-term credit market. In the base scenario \(\beta = 0.05\), the repo layer features systemic impacts up to an order of magnitude larger, with some nodes accounting for up to 10\% of the repo market in the event of default. This gap narrows as \(\beta\) increases, with \(\beta = 0.2\) yielding more similar DebtRank values across the two layers.

The non-linear sensitivity to liquidity buffers is highlighted in the last column of Figure~\ref{DR_liq}. The increase in DebtRank from \(\beta = 0.1\) to \(0.2\) is substantially larger than the increase from \(\beta = 0.05\) to \(0.1\), for both layers. This illustrates how tighter liquidity conditions can amplify roll-over risk and contagion potential. Finally, while the repo market exhibits high systemic impact under this calibration, it is also highly collateralized; in practice, collateral liquidation can mitigate liquidity shortfalls and reduce propagation.\footnote{An extension of the current framework would be to include collateral information in future work.}

We can also examine whether non-linearities exist across layers: when propagating an exogenous shock through multiple layers simultaneously, does systemic impact exceed what would be implied by treating layers independently?

\begin{figure}[!htbp]
\centering

\begin{subfigure}[t]{0.48\textwidth}
  \centering
  \includegraphics[width=\linewidth]{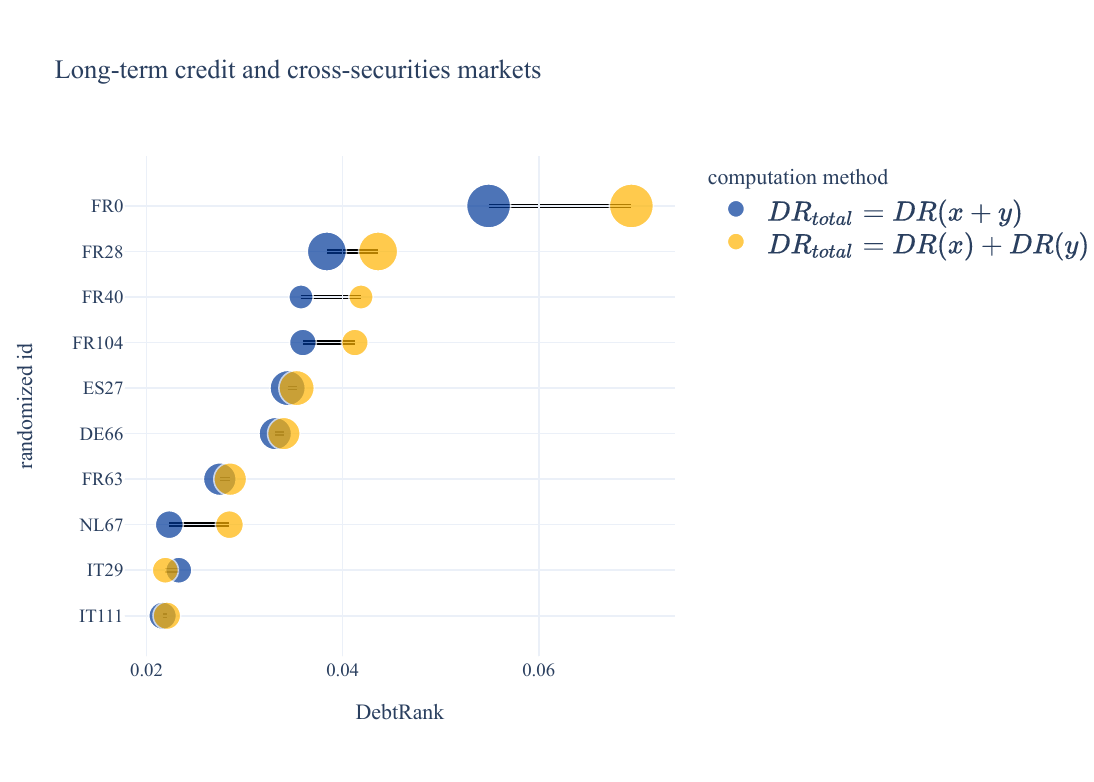}
  \caption{Long-term credit market and cross-securities market.}
  \label{fig:DR_superposition_counterparty}
\end{subfigure}\hfill
\begin{subfigure}[t]{0.48\textwidth}
  \centering
  \includegraphics[width=\linewidth]{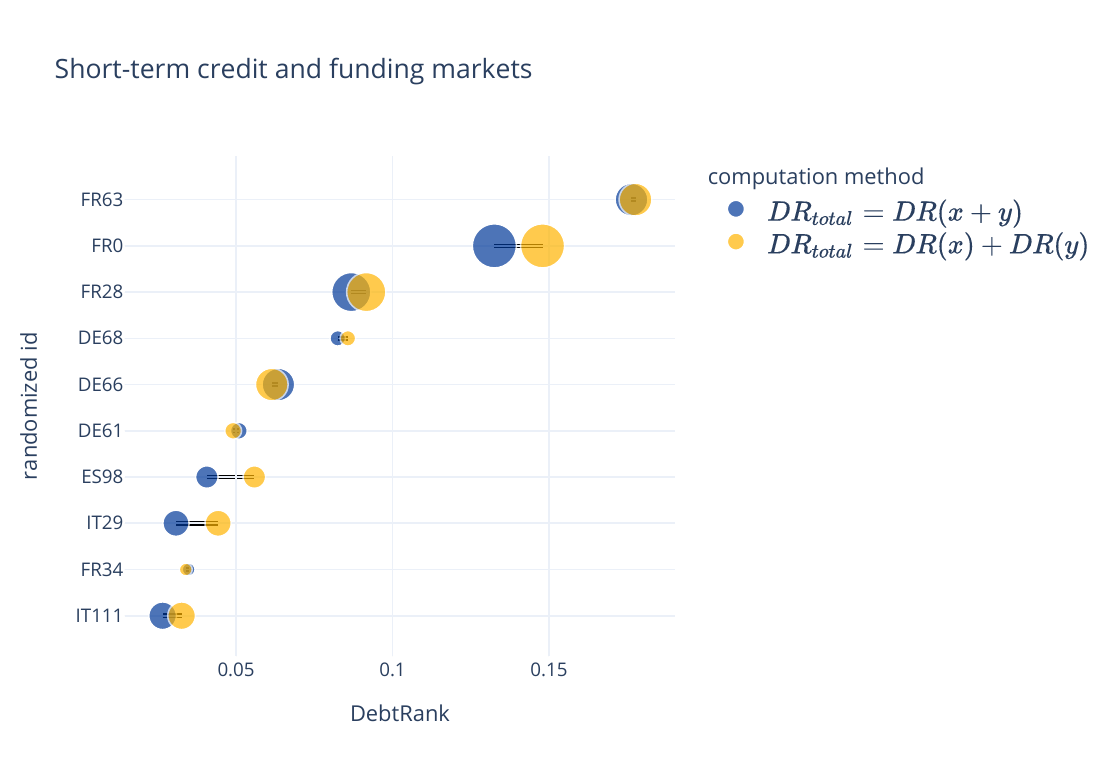}
  \caption{Short-term credit market and funding market.}
  \label{fig:DR_superposition_liq}
\end{subfigure}

\caption{
Left panel depicts DebtRank values when shocking simultaneously the long-term credit market and the cross-securities market; right panel focuses on the short-term credit market and funding market. The x-axis depicts DebtRank, whereas the y-axis depicts the top 10 banks with the largest systemic impact. The blue markers are DebtRank values when computed on the aggregated layers, whereas yellow markers are based on a linear superposition of the DebtRank values computed for the layers separately. The figure shows results for \(\beta = 0.1\), although different values yield the same qualitative results. The top 10 banks with highest DebtRank values are determined by taking the average results of the two computation methods.
}
\label{fig:DR_superposition}
\end{figure}

Figure~\ref{fig:DR_superposition} reports this thought experiment. The left panel considers shocks propagated jointly through the long-term credit and cross-securities layers, whereas the right panel focuses on the short-term credit and repo layers.\footnote{The figure shows results for \(\beta = 0.1\), although different values yield the same qualitative results.} The x-axis depicts DebtRank, and the y-axis lists the top 10 banks by systemic impact. Blue markers correspond to DebtRank computed on the aggregated (combined) layer, while yellow markers represent a linear superposition of layer-by-layer DebtRank values.\footnote{The top 10 banks with highest DebtRank values are determined by taking the average results of the two computation methods.} Differences between blue and yellow markers capture interaction effects induced by aggregation. In the left panel, linear superposition tends to overstate systemic impact relative to the direct aggregated computation. In the right panel, differences are smaller and less systematic. Overall, with the exception of a few highly systemic nodes for which aggregation induces more noticeable deviations, the proximity of the two computations suggests limited additional non-linearities from simple layer aggregation in this setting.

\subsection{Micro-structural agent-based contagion framework}\label{agent_mod} 

Our second analytical use case leverages an agent-based modelling framework based on the micro-structural multi-layer approach of~\cite{montagna2016multi}, extended in~\cite{montagna2020origin}. In contrast to the original paper, we use real-world exposures from our multilayer network rather than simulated edges between nodes. This allows us to evaluate a standard micro-structural propagation mechanism in a setting where network topology and exposure magnitudes are disciplined by granular data.

We assume that a shock first passes through the long-term interbank market, where banking groups borrow and lend to each other on a long-term basis and issue and invest in shares and obligations to each other. This market is primarily embedded with counterparty risk, i.e., the risk that a debtor cannot meet its obligations, forcing creditors to book losses. Similarly, if an investee goes into default, its investors book corresponding losses.
Outcomes from the long-term market may induce a chain reaction in the short-term liquidity market where banking groups use short-term debt to finance balance-sheet activities; this market is embedded with funding and roll-over risk.
When a creditor hoards liquidity, its debtors may struggle to refinance and meet repayments. Lastly, contagion can propagate through the external securities market where securities are marked-to-market and banking groups hold potentially overlapping portfolios. This market is embedded with market risk: liquidation by one node can depress prices and deteriorate the balance sheets of other holders, potentially generating a fire-sale spiral.

The agent-based framework allows us to assess systemic importance in terms of how the default of a banking group influences the system through these channels. We take a similar approach as for the DebtRank analysis and, for each node \(n_{i}\), initiate a simulation in which $n_i$ goes into default due to an exogenous shock.
The default corresponds to a full depletion of $n_i$'s cash and equity balances, leading to the violation of regulatory requirements and the freezing of its securities portfolio. Creditors and investors of \(n_{i}\) on the long-term market book losses accordingly; if this reduces their capital ratios below a threshold they begin hoarding liquidity on the short-term market, propagating the initial shock further. We track (i) the number of additional defaults induced by \(n_{i}\) and (ii) the cumulative capital base of defaulting banks until convergence (i.e., when no additional defaults occur). We repeat this exercise with each node as the initially defaulting entity.
We refer to Appendix~\ref{sec:appendix} for a more detailed exposition of the model and its dynamics.

\begin{figure}[!htbp]
\centering

\begin{subfigure}[t]{0.49\textwidth}
  \centering
  \includegraphics[width=\linewidth]{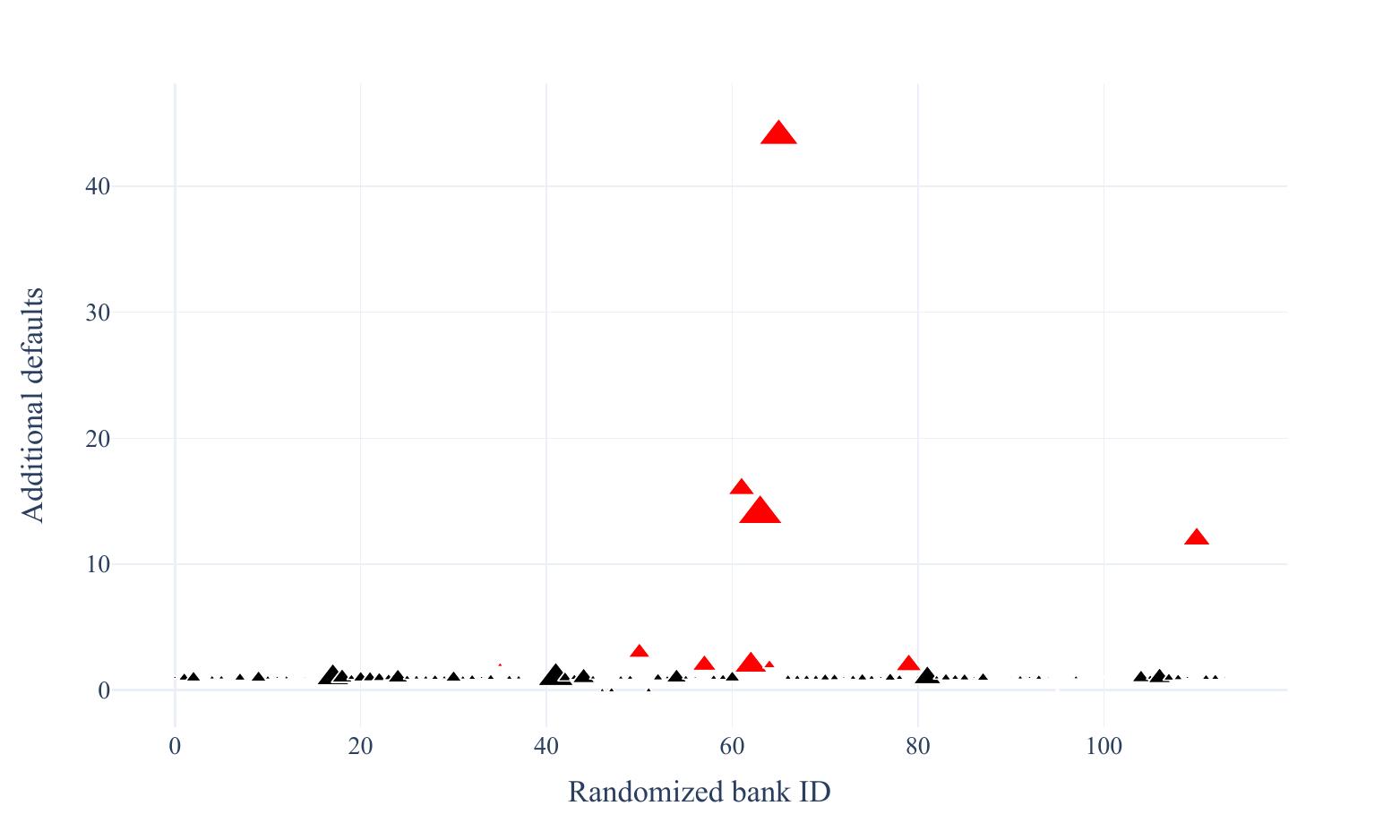}
  \caption{Additional number of defaults triggered by an initial default.}
  \label{fig:shock_prop_additional_defaults}
\end{subfigure}\hfill
\begin{subfigure}[t]{0.49\textwidth}
  \centering
  \includegraphics[width=\linewidth]{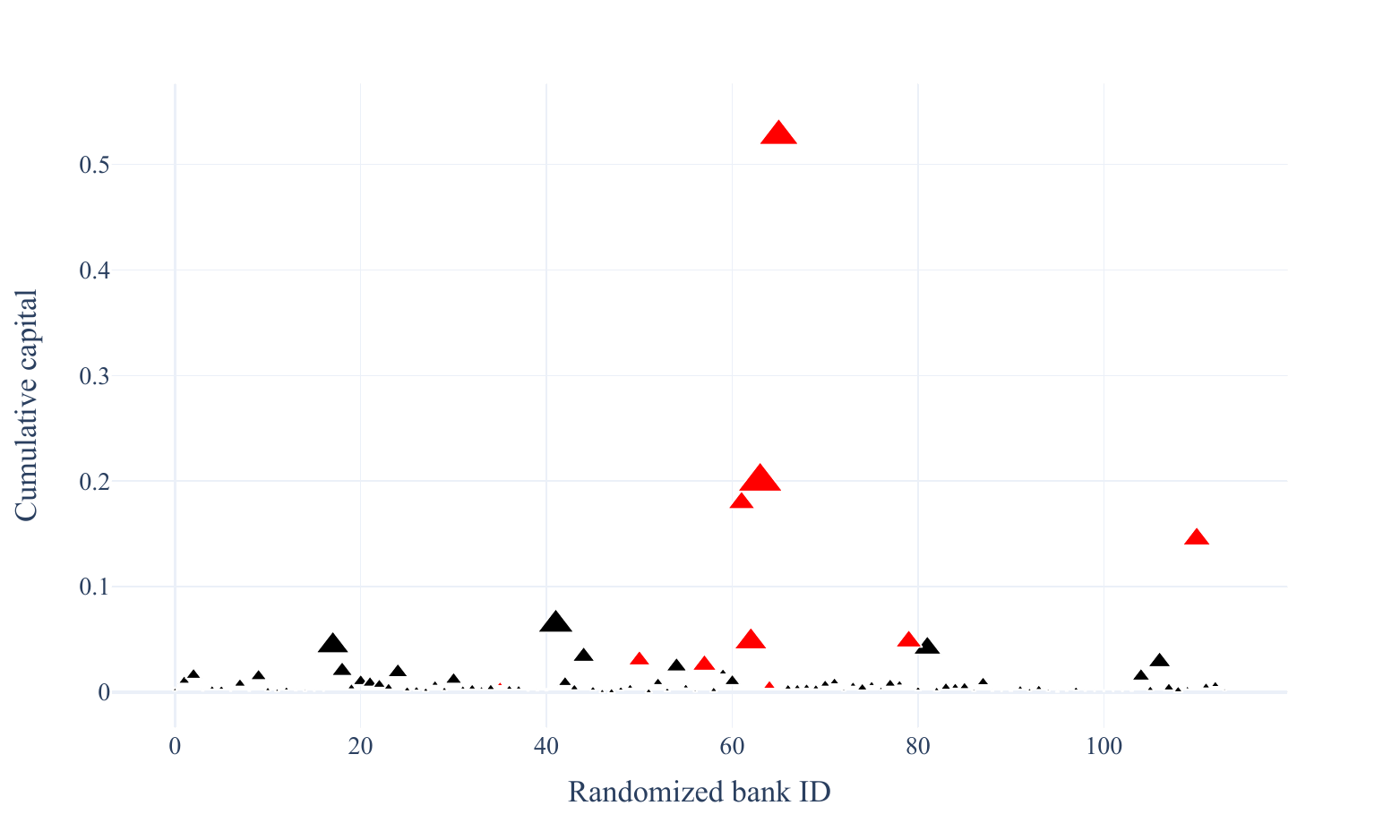}
  \caption{Cumulative capital base of defaulting institutions.}
  \label{fig:shock_prop_cumulative_capital}
\end{subfigure}

\caption{
Panel (a) reports the additional number of defaults induced by an initially defaulting institution \(n_i\).
Panel (b) reports the cumulative capital base of defaulting institutions, including \(n_i\).
Marker size is proportional to total assets. Institutions that induce contagion effects are shown in red; institutions with no contagion effects are shown in black.
The x-axis reports anonymized bank identifiers.
Risk weights are set to \(w^{\mathrm{ib}}=0.2\) for interbank exposures and \(w^{\mathrm{es}}=0.1\) for external securities.
Price-impact weights are set to \(\alpha=0.2\).
The liquidity-buffer scaler is \(\beta=0.05\), and the minimum capital ratio is \(\overline{\gamma}=0.10\).
}
\label{shock_prop}
\end{figure}

In Figure~\ref{shock_prop}, the left panel reports the number of additional defaults, while the right panel reports the cumulated capital base of defaulting entities as a fraction of total capital in the system. Most institutions do not generate contagion (zero additional defaults). For a small subset, however, initial defaults trigger cascades: some nodes induce up to two additional defaults, while a smaller set generates cascades of 10 to 15 additional defaults. For one entity, default triggers a long chain across multiple cycles, leading to 43 additional defaults, indicating far-reaching propagation that reaches seemingly distant nodes.

All entities that generate contagion effects originate from only five countries, suggesting national clustering effects. While several contagion-inducing institutions are large, we also observe smaller institutions that induce propagation, and conversely large institutions that do not. This underscores that size alone has limited explanatory power for systemic impact, consistent with theoretical results obtained on artificial networks~\parencite{glasserman2015likely}.
The bottom panel shows that the cumulated capital base remains below 10\% for most cascades, but four nodes induce spirals exceeding 10\%, reaching more than 50\% for the most impactful entity.

\begin{figure}[!htbp]
\centering

\begin{subfigure}[t]{0.49\textwidth}
  \centering
  \includegraphics[width=\linewidth]{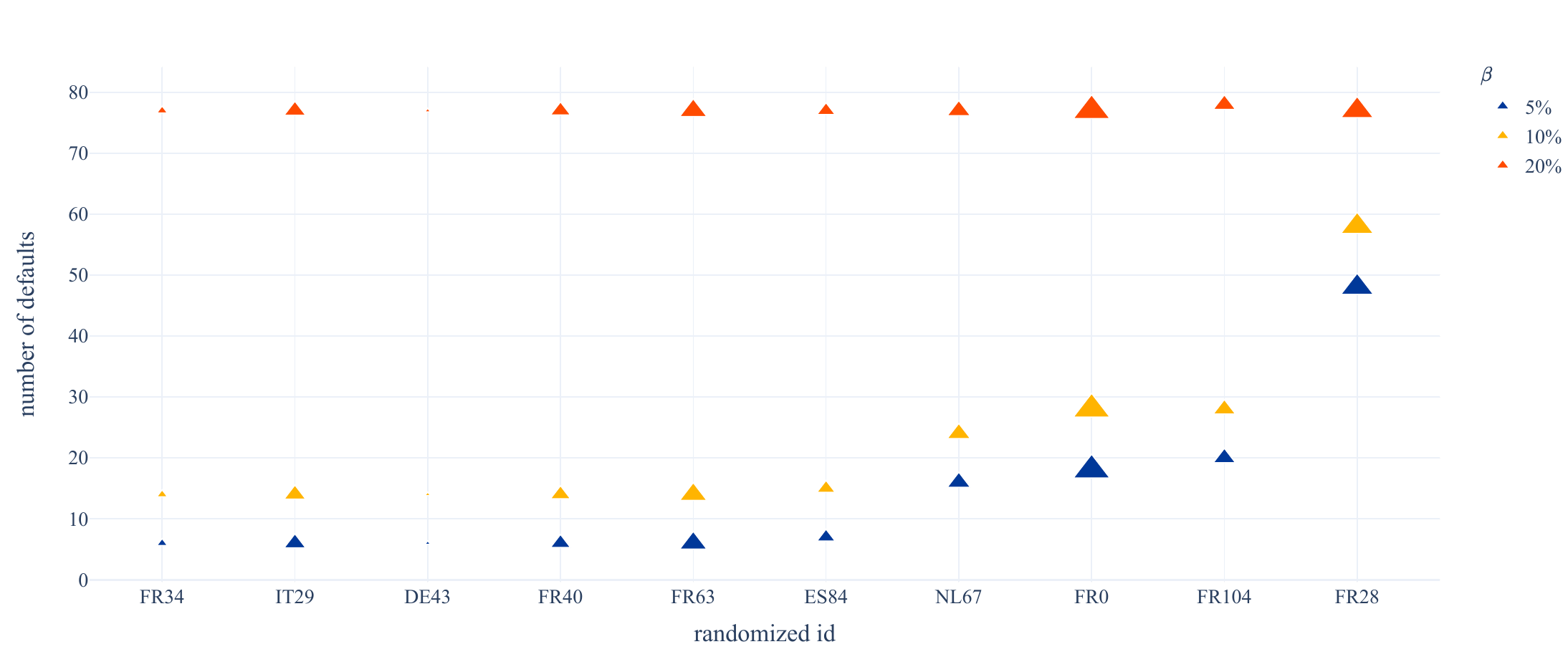}
  \caption{Additional defaults induced by the initial default.}
  \label{fig:beta_dyn_defaults}
\end{subfigure}\hfill
\begin{subfigure}[t]{0.49\textwidth}
  \centering
  \includegraphics[width=\linewidth]{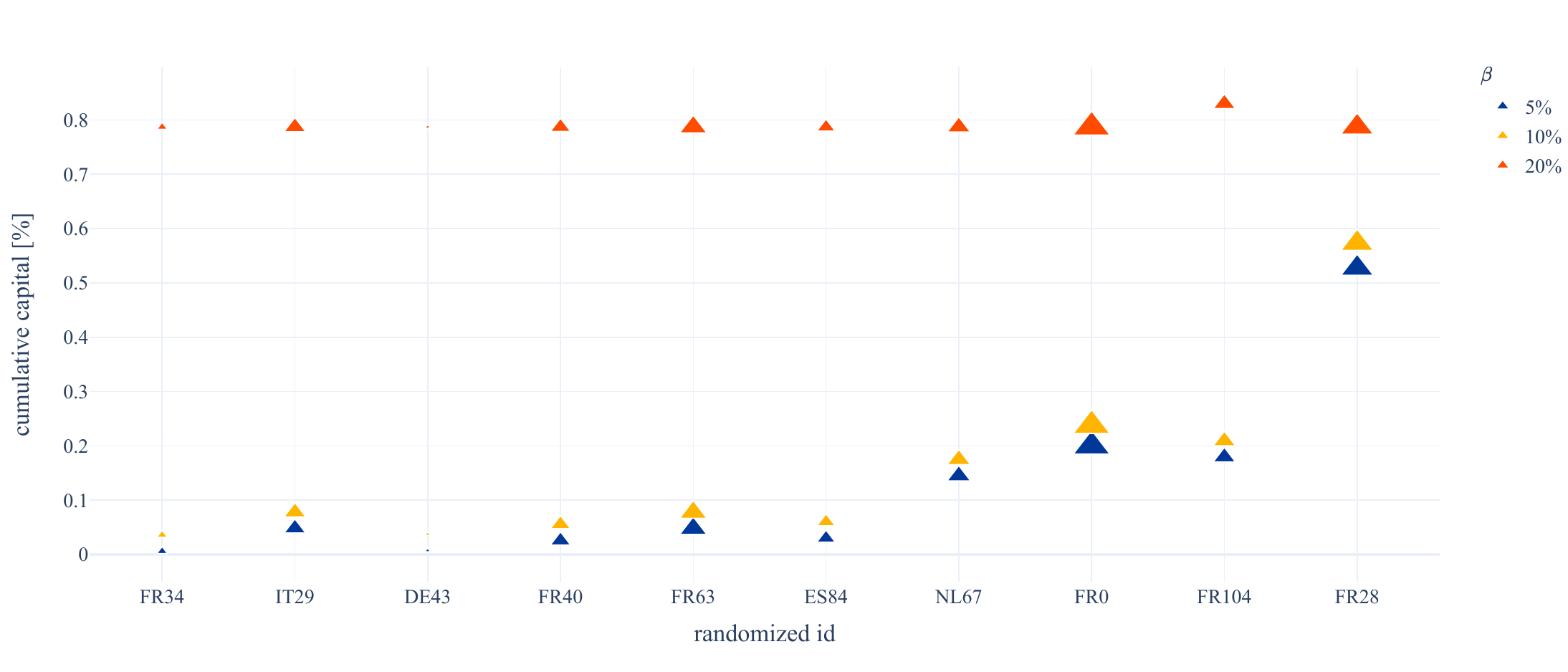}
  \caption{Cumulative capital base of defaulting institutions.}
  \label{fig:beta_dyn_capital}
\end{subfigure}

\caption{
Panel (a) reports the additional number of defaults induced by an initially defaulting institution \(n_i\).
Panel (b) reports the cumulative capital base of defaulting institutions, including \(n_i\).
Each initially defaulting institution is selected among the most systemic nodes in the network.
The x-axis reports anonymized bank identifiers; the first two characters indicate the country of headquarters.
Risk weights are set to \(w^{\mathrm{ib}}=0.2\) for interbank exposures and \(w^{\mathrm{es}}=0.1\) for external securities.
Price-impact weights are set to \(\alpha=0.2\).
The liquidity-buffer scaler varies over \(\beta \in \{0.05, 0.10, 0.20\}\), and the minimum capital ratio is \(\overline{\gamma}=0.10\).
}
\label{beta_dyn}
\end{figure}

Figure~\ref{beta_dyn} zooms into the 10 most systemically important institutions and studies sensitivity to the liquidity buffer scaler $\beta$. A higher $\beta$ corresponds to greater net outflows of short-term funds and thus higher liquidity risk. Increasing $\beta$ from 0.05 to 0.1 produces stronger contagion: the number of additional defaults rises on average by 8 and the cumulative defaulted capital base by 3.2\% (a relative increase of 27\%). Increasing $\beta$ further to 0.2 yields an extreme increase, with all 10 institutions producing large cascades (around 75 additional defaults). The uniformity at $\beta=0.2$ suggests an upper bound on default counts that is largely determined by network topology. A similar non-linear pattern emerges for the cumulative capital base, reaching around 80\% (an average relative increase of roughly 500\%).
These findings align with the DebtRank sensitivity analysis and highlight the importance of maintaining adequate liquidity buffers to mitigate contagion when liquidity hoarding is present.

\FloatBarrier
\section{Summary and conclusions}
\label{sec:summary}

The Great Financial Crisis brought to the fore the relevance of interconnectedness in interbank markets and the need for detailed data to understand the relationships that link credit institutions. Subsequent episodes, including the Great Recession and the Covid-19 shock, further underscored that macro-financial outcomes can be shaped by propagation mechanisms operating through multi-faceted and far-reaching linkages among financial entities.

This paper advances an empirically grounded, system-level representation of euro area banking interconnectedness. We integrate multiple granular supervisory and statistical collections into a multilayer network of significant banking groups, enabling a layer-consistent view of exposures across distinct transmission channels. The multilayer perspective reveals pronounced heterogeneity in both topology and institutional roles across markets; heterogeneity that can be muted or misrepresented when exposures are collapsed into a single aggregate network.

Building on this integrated measurement framework, we extend leading network-based systemic-risk approaches to a setting in which exposures are observed jointly across markets and can be analysed coherently at the layer level. In particular, we adapt a propagation-based systemic-importance measure and a micro-structural agent-based model to operate on the empirically constructed multilayer network. The propagation analysis identifies channel-specific systemic importance and documents a strong sensitivity of short-term market contagion to liquidity buffers, with the ranking of influential institutions varying across layers. The agent-based framework, by explicitly coupling balance-sheet adjustments, roll-over decisions, and fire-sale spillovers across market segments, generates sizeable default cascades for a small subset of institutions and reinforces that systemic impact is shaped not only by size, but also by network position and market-specific connectivity.

These results underscore the value of assessing resilience in a layer-aware manner when the same institutions participate in multiple market segments with different micro-structural features. A multilayer, empirically calibrated view of exposures can inform policy discussions on the design of flexible and targeted macro- and micro-prudential instruments, and provides a concrete empirical basis for evaluating multi-channel contagion mechanisms in stress-testing exercises.

\printbibliography[heading=bibintoc, title=\ebibname]


\appendix
\section{Agent-based modelling framework}\label{sec:appendix}

The main model deployed in this paper is heavily based on the microstructural multi-layered approach introduced in \cite{montagna2016multi}. In contrast to the original paper, we do not rely on simulated edges between nodes, but utilize our multi-layer network constructed of real-world granular data. The agent-based model allows one to traverse an initial exogenous default shock throughout the different layers of the multi-layer network, considering the distinct characteristics and embedded risks within the layers.

Each node \(n_i\) is characterized by a simplified balance sheet (Table~\ref{tab:bs_simplified}).

\begin{table}[!htbp]
\centering
\caption{Simplified balance sheet of banking groups}
\label{tab:bs_simplified}
\begin{tabular}{@{}ll@{}}
\toprule
Assets & Liabilities \\
\midrule
\(c_i\): cash & \(d_i\): deposits \\
\(e_i\): external securities & \(eq_i\): equity \\
\(u_i^{a}\): cross-securities holdings & \(u_i^{l}\): cross-securities issuance \\
\(l_i^{l}\): long-term loans & \(b_i^{l}\): long-term borrowings \\
\(l_i^{s}\): short-term loans & \(b_i^{s}\): short-term borrowings \\
\(h_i^{a}\): repurchase agreements & \(h_i^{l}\): short-term funding \\
\(o_i^{a}\): other assets & \(o_i^{l}\): other liabilities \\
\bottomrule
\end{tabular}
\end{table}

Next, we define the propagation logic of the model. We assume that nodes
can be characterized by a state variable
\(\mathbf{\phi} \in \left\{ 0,1,2 \right\}^{n}\):

\[
\phi_{i} = \begin{cases}
    2, & \textit{ if $n_i$ in default}\\
    1, & \textit{ if $n_i$ in distress}\\
    0, & \textit{ otherwise}
            \end{cases}
\]

We further assume that nodes will take the necessary measures possible
to avoid going into distress or default.

There are two main channels that impact the state of a node. The first
channel is determined by the capital of the nodes. Nodes go into a
distressed state if at any point in time we observe:

\[\gamma = \frac{\text{Eq}}{w^{b} \cdot \left( l^{l} + l^{s} + u^{a} \right) + S\left( w^{s} \cdot P \right) + C^{\text{te}}} < \ \overline{\gamma}\]

With \(w^{b} \in \mathbb{R}^{n},\ w^{s} \in \mathbb{R}^{m}\)
respectively the risk-weights for interbank assets and external
securities. \(S \in \mathbb{R}^{n \times m}\ \) is the holding matrix of securities
  with \(s_{i\mu}\) the amount that node \(n_{i}\) is holding of
  security \(\mu\). $P \in \mathbb{R}^{m}$ is the price vector of securities with $p_{\mu}$ the price of security \(\mu \in M\) and $C^{\text{te}}$ a constant representing risk-weighted
assets that are not explicitly modelled. \(\gamma \in \mathbb{R}^{n}\)
is thus a proxy for the risk-weighted capital ratio of banks and needs
to be above a minimum regulatory lower bound \(\overline{\gamma}\).

The second channel that might impact the state of a node is determined
by the liquidity position of the node. Nodes go into a distressed state
if we observe at any point in time:

\[c < \beta\left( d + b^{s} + h^{l} \right)\]

With \(\beta\) a liquidity buffer scaler.

We further assume that a propagation cycle coincides with the longest
maturity of short-term debt, that is during the shock propagation nodes
need to decide how much of their short-term debt to roll-over to their
current debtors. In case a node decides to withdraw its short-term funds
from the market, its debtors might be facing difficulties in actually
repaying the debt. This gives rise to an additional liquidity channel
that impacts the state of a bank by the end of a cycle. This is encoded
as:

\[p < \overline{p}\]

With \(\overline{p} \in \mathbb{R}^{n}\) the proportion of short-term
debt that banks need to repay from their total short-term borrowings,
and \(p \in \mathbb{R}^{n}\) the clearing vector that clears the
short-term interbank market during a certain propagation cycle. Note the
correspondence with an~\cite{eisenberg2001systemic} system.

If a node \(n_{i}\) admits to any one of the three above violations it is
put in a distressed state. Node \(n_{i}\) will then try to rectify its
position by (i) withdrawing liquidity from the short-term interbank
market and, if further needed, by (ii) selling off (a portion) of its
external securities. If after trying to do so, \(n_{i}\) is still in a
distressed state it will be put in a default state. When a node \(n_{i}\)
goes into default its securities portfolio gets frozen by external
authorities until the markets are stable again. This to avoid inducing
further fire sales into the system. Next, the remaining nodes will have
to book their losses on the long-term interbank market in case they had
exposures against the defaulted nodes. This will induce a second cycle
in the shock propagation mechanism and will continue to do so until the
system converges, i.e. when no additional defaults are observed at the
end of a propagation cycle.

\subsection{The long-term interbank market}

The long-term interbank market is constructed based on two layers of the multi-layer network, the long-term credit, and the cross-securities layers. At the beginning of each propagation cycle, nodes book losses from the long-term interbank market, if any, due to the bankruptcy of their debtors/investee’s in the previous period. These losses affect the capital of the node and therefore can lead to the violation of their capital ratio constraint. To be more precise, and in line with~\cite{ContMoussaSantos}, when a node $n_i$ defaults, it leads to the immediate write-down in value of all its liabilities to its creditors/investors. These losses are then imputed to the Tier 1 capital base (represented by the $Eq \in \mathbb{R}^{n}$ vector) of the creditors, leading to a deterioration of the risk-weighted capital ratio of these creditors/investors. In case the updated capital ratio is lower than the regulatory required lower bound $\bar{\gamma}$, the creditors and investors become distressed and will try to gain liquidity from the short-term interbank market to rebalance their capital ratio.

\subsection{The short-term interbank market}

If one of the nodes constraints is violated, it will need to react
adequately to avoid going into default. The first step to undertake is
to reduce its short-term interbank market exposures. We assume that each
cycle in the model coincides with the longest maturity level of
short-term debt, that is at the end of each cycle nodes need to decide
on the amount of their short-term debt they want to roll-over to their
debtors for next period.

This endogenous decision is based on two factors: (i) the need to make sure a node's constraints are satisfied and (ii) the fact that a nodes own funding from other creditors is being reduced, forcing it to withdraw more money from its debtors.

This endogenous decision making can be formulated by the following
mapping:

\[f \cdot l = min\{ r + \max\left\{ \left( W_{\text{st}}^{T}f - c_{\text{buf}} \right),\ 0 \right\},\ l\}\]

With,

\begin{itemize}
\item
  \(f \in \left\lbrack 0,1 \right\rbrack^{n}\) the fraction of
  short-term debt that will not be rolled-over
\item
  \(l = l^{s} + u^{a}\), the total short-term assets available to
  roll-over,
\item
  \(W_{\text{st}}^{T} = W_{short\_ term\_ credit}^{T} + W_{short\_ term\_ funding}^{T}\),
  the sum of the adjacency matrices derived from the short-term credit
  and funding layers. \(\left( w_{\text{st}}^{T} \right)_{\text{ij}}\)
  thus represents the total short-term debt that \(n_{i}\) has received from
  \(j\), and \(W_{\text{st}}^{T}f\) depicts then the amount of debt that
  needs to be repaid to nodes
\item
  \(c_{\text{buf}} = c - \beta(d + b^{s} + h^{l} - W_{\text{st}}^{T}f)\)
  represents the cash level above the liquidity requirement that can
  readily be used to pay back nodes
\item
  \(r \in \mathbb{R}^{n}\) a vector denoting the amount of funds each
  bank wants to withdraw to satisfy their constraints. We can decompose
  \(r\) in two additive components being:

  \begin{itemize}
  \item
    \(r_{\text{liq}} = \min\left\{ \max\left\{ \beta\left( d + b^{s} + h^{l} - W_{\text{st}}^{T} \cdot f \right) - c,\ 0 \right\},\ \ l \right\}\)
  \item
    \(r_{\text{cap}} = min\{ max\{\frac{\overline{\gamma}(C^{\text{te}} + S\left( w^{s} \cdot P \right) + \ \overline{\gamma}w^{b}\left( l^{l} + l - r_{\text{liq}} \right) - Eq}{\overline{\gamma}w^{b}},\ 0\},\ l - r_{\text{liq}}\}\)
  \end{itemize}
\end{itemize}

with \(r_{\text{liq}}\) taking care of liquidity requirements and
\(r_{\text{cap}}\) taking care of capital requirements. The above loop
describes how a node withdraws funds from the short-term market only if
it has problems fulfilling its constraints and in case other nodes
decide to withdraw their funds that are deposited with the node.

Note the similarities between the above mapping and the mapping
introduced by ~\cite{eisenberg2001systemic}. We can leverage the same
mathematical tools as used in ~\cite{eisenberg2001systemic} to solve the
above decision-making process. To do so we need to rewrite the above,
noting that the mapping is homogenous of degree one, that is:

\[f = min\{\frac{r + \max\left\{ \left( W_{\text{st}}^{T}f - c_{\text{buf}} \right),\ 0 \right\}}{l},\ 1\}\]

Allowing us to reformulate the problem as a fixed-point problem of the
mapping:

\[\Phi(f):\left\lbrack 0,1 \right\rbrack^{n} \rightarrow \left\lbrack 0,1 \right\rbrack^{n} = \ min\{\frac{r + \max\left\{ \left( W_{\text{st}}^{T}f - c_{\text{buf}} \right),0 \right\}}{l},1\}\ \]

Solving the fixed-point problem yields then the short-term equilibrium
state of the short-term interbank market during a propagation cycle.

\subsection{The (external) securities market}

Once banks decide about how much to withdraw from the interbank market,
their counterparts might need to sell-off portions of their securities
portfolio, above their cash and short-term funds, in order to
successfully repay their creditors. The securities market is constructed
based on the external-securities layer of the multi-layer network.

To properly model the securities market, we need to know (i)
the prices of the securities in our universe, (ii) the amount that each
node \(n_{i}\) is holding of security \(\mu\) and (iii) the way that prices
change through time (which will be endogenously determined). We
therefore use:

\begin{itemize}
\item
  The price vector of securities \(P \in \mathbb{R}^{m}\) with
  \(p_{\mu}\) the initial price of security \(\mu \in M \). $M$ represents the set of all securities in our universe, with $m$ the total number of securities as noted before.
\item
  The holding matrix of securities \(S \in \mathbb{R}^{n \times m}\ \)
  with \(s_{i\mu}\) the amount that node \(n_{i}\) is holding of
  security \(\mu\) with $n_i \in N$ and
  $\mu \in M$.
\item
  The total market value of \(n_{i}\)'s external securities portfolio can
  then be simply represented by:
\end{itemize}

\[\ E = SP\]

With \(e_{i}\ \) the market value of \(n_i\)'s portfolio.

\begin{itemize}
\item
  Lastly, the sell-off matrix \(Z \in \mathbb{R}^{n \times m}\) with
  \(z_{i\mu}\) the amount that node \(n_{i}\) is selling of security
  \(\mu\) at any given moment in time. Note that
  \(Z \in \lbrack 0,\ S\rbrack\).
\end{itemize}

We furthermore assume that the price dynamics of the securities are
governed by the following process:

\[P^{t + 1} = P^{t}e^{- \alpha \cdot \delta}\]

With
\begin{itemize}
    \item \(\alpha \in \left\lbrack 0,1 \right\rbrack^{m}\) a vector of constants
representing the narrowness of the market for the securities,
    \item \(\delta =\)
\(\frac{\mathbf{1}_{\mathbf{n}}Z}{\mathbf{1}_{\mathbf{n}}S} \in \mathbb{R}^{m}\),
with \(\delta_{\mu}\) the proportion of security \(\mu\) that has been
sold off by the nodes in the network relative to the total available
amount of that security in the network
\end{itemize}

Note how the price dynamics assumes that prices cannot increase from one
period to the other, as we model an extreme tail scenario over a short
total timeframe.

We can look at the proportion being sold from two perspectives, (i) the
proportion being sold relative to the initial total holding amount
within the interbank network or (ii) the proportion being sold relative
to the total holding amount available within the interbank as of last
period. (ii) is a more sophisticated approach that induces a memory
effect in the price dynamics: The more of a security has been sold
during a fire sale spiral the harder the price will decrease for each
additional sell that happens.

\begin{figure}[!htbp]
\centering
\includegraphics[width=0.9\textwidth]{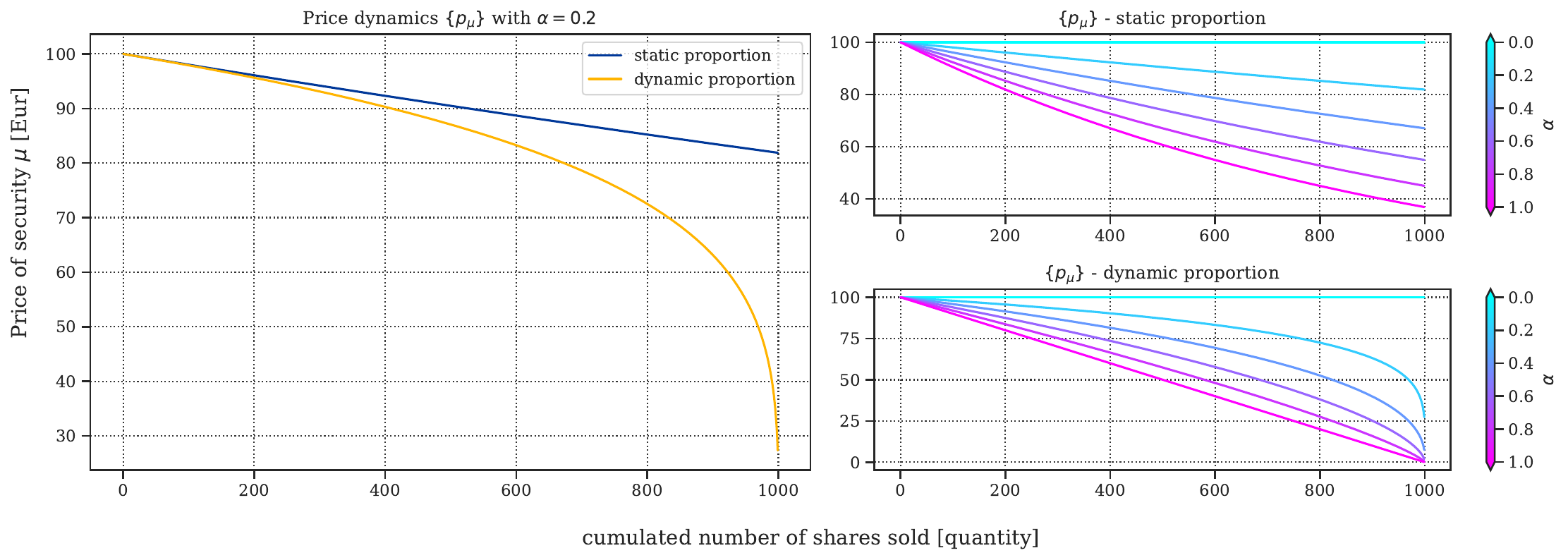}
\caption{
Left panel: price dynamics with impact parameter \(\alpha_{\mu} = 0.2\),
for both static proportion computation (blue) and dynamic proportion
computation (yellow). Right upper panel: price dynamics of static
proportion-based computation for various \(\alpha_{\mu}\) values. Right
bottom panel: price dynamics of dynamic proportion-based computation for
various \(\alpha_{\mu}\) values. X-axis depict the quantity of \(\mu\)
sold, y-axis shows the resulting price of \(\mu\).
Left panel shows a comparison of the endogenous price mechanism
\(P^{t + 1} = P^{t}e^{- \alpha \cdot \delta} \)when using a static vs.
dynamic proportion-based computation of the amount of securities sold.
The dynamic based computation shows a sharper decrease of the price of a
security due to memory effects. The right panels show price
depreciations for different values of \(\alpha\) for the static based
method (upper) and dynamic based method (lower). The dynamic proportion
can lead to extreme price depreciation when the market is very narrow.
}\label{price_dyn}
\end{figure}

Fig~\ref{price_dyn} illustrates the two different perspectives. The
exponential decrease of (ii) (depicted as the yellow graph in the left
figure) shows the memory effect. When \(\alpha_{\mu}\) is high we see
that the price of a security \(\mu\) goes down to a lower bound that is
greater than 0 for (i) whereas the price goes to 0 for (ii) (see upper
and bottom panel of the right side of the figure). This edge case can
for example represent a highly illiquid security such as an RMBS product
during the Great Financial Crisis. To adequately model such an
environment we take perspective (ii) moving forward. ~

Again, we rely on ~\cite{eisenberg2001systemic} to find the optimal \(Z\)
that helps clear the system:

\[p = min\{{c + \ \Pi}^{T}p + ZP,\ \overline{p}\}\]

With,

\begin{itemize}
\item
  \(\overline{p} = W_{\text{st}}^{T}f \in \mathbb{R}^{n}\) the
  proportion of short-term debt that nodes need to repay from their
  total short-term borrowings based on the roll-over decisions made in
  the short-term interbank market.
\item
  \(p \in \mathbb{R}^{n}\) the clearing vector that clears the
  short-term interbank market within a propagation cycle

\item \(\Pi \in {\lbrack 0,1\rbrack}^{n \times n}\) the relative liabilities
  matrix with \[
  \pi_{\text{ij}} =
                \begin{cases}
                    \frac{\left( w_{\text{st}}^{T} \right)_{\text{ij}} \cdot f_{i}}{\overline{p}}, & \textit{ if $\overline{p}>0$} \\
                    0, & \textit{otherwise}
                \end{cases}
                \]
    the liability that node \(n_{i}\) has against
  \(j\) relative to \(n_i\)'s total obligations.

\item
  \(ZP \in \mathbb{R}^{n \times m}\) gives the liquidity gained from
  selling securities, with the price vector \(P\) in turn dependent on
  how much securities actually being sold, as driven by the endogenous
  price dynamics of the system.
\end{itemize}

We further assume that a node \(n_{i}\) in liquidity need sells securities
in its portfolio proportional to the market value of its holdings while taking into account the risk-weights of the securities. That is,
ceteris paribus, securities with higher risk weights are sold in
larger quantities than securities with lower risk weights. The
underlying logic is that banks attempt to be efficient in restoring their
risk-weighted capital ratios. We obtain the proportional weight matrix of portfolios as,

\[\eta = \frac{S \cdot (w^{s} \cdot P)}{(S \cdot \lbrack w^{s} \cdot P\rbrack)1_{n}} \in \mathbb{R}^{n \times m}\]

We can then decompose the sell-off matrix \(Z\) into three additive
components, one for each of the requirements banks must satisfy:

\begin{enumerate}
\item
  Banks must ensure sufficient liquidity to repay creditors in the short-term interbank market, based on roll-over decisions:

\[\text{Li}q_{\text{need}}^{\text{interbank}} = \max\left\{ \overline{p} - \left( c_{\text{buf}} + \Pi^{T}p \right),\ 0 \right\} \in \mathbb{R}^{n}\]

The amount to be sold of each security is then,

\[Z^{\text{interbank}} = min\{\frac{\left( \text{Li}q_{\text{need}}^{\text{interbank}} \right)^{T} \cdot \eta}{P},\ S\}\]

\item
  Banks must also satisfy the liquidity requirement,

\[\text{Li}q_{\text{need}}^{\text{liquidity}} = max\{\beta\left( d + b^{s} + h^{l} - W_{\text{st}}^{T}f \right) - c,0\}\]

The amount to be sold of each security is then,

\[Z^{\text{liquidity}} = \min\left\{ \frac{\left( \text{Li}q_{\text{need}}^{\text{liquidity}} \right)^{T} \cdot \eta}{P},\ S - Z^{\text{interbank}} \right\}\]

\item
  Lastly, banks must maintain their capital requirement, recognizing that fire-sale price effects can deteriorate balance sheets:
\end{enumerate}

\[\text{Li}q_{need\_ rw}^{\text{capital}} = \max\left\{ w^{b} \cdot A + B\left( w^{s} \cdot P \right) + C^{\text{te}} - \frac{\text{Eq}}{\overline{\gamma}}\ ,0 \right\}\]

With,

\begin{itemize}
\item
  \(A = \ l^{l} + l^{s} + h^{a} - \left( r_{\text{liq}} + r_{\text{cap}} \right)\)
\item
  \(B = \ S - (Z^{\text{interbank}} + Z^{\text{liquidity}})\)
\end{itemize}

Cancelling out the risk-adjusted measure yields liquidity needs in euros:

\[\text{Li}q_{\text{need}}^{\text{capital}} = \left( \frac{\text{Li}q_{need\_ rw}^{\text{capital}} \cdot \eta}{w^{s}} \right)1_{m} \in \mathbb{R}^{n}\]

Raising \(\text{Li}q_{\text{need}}^{\text{capital}}\) ensures that the
capital ratio is at least \(\overline{\gamma}\).

The amount to be sold of each security is then,

\[Z^{\text{capital}} = \min\left\{ \frac{\left( \text{Li}q_{\text{need}}^{\text{capital}} \right)^{T} \cdot \eta}{P},\ S - (Z^{\text{interbank}} + Z^{\text{liquidity}}) \right\}\]

These intermediary results yield
\(Z = Z^{\text{interbank}} + Z^{\text{liquidity}} + Z^{\text{capital}} \in \left\lbrack 0,\ S \right\rbrack^{n \times m}\),
the sell-off matrix of the system. Solving the fixed-point
problem of the mapping

\[\xi\left( p \right):\left\lbrack 0,\ \overline{p} \right\rbrack^{n} \rightarrow \left\lbrack 0,\ \overline{p} \right\rbrack^{n}:\ min\{{c + \ \Pi}^{T}p + ZP,\ \overline{p}\}\ \]

produces the clearing vector \(p\) and, as a byproduct, the sell-off matrix \(Z\) and updated price vector \(P\). The mechanism can be interpreted as follows: liquidity needs in the interbank market induce security sales, which depress prices; price declines erode capital ratios, which in turn can raise liquidity needs and further sales. Once \(p\) is sufficient to clear obligations, \(Z\) and \(P\) stabilize and the system reaches an equilibrium for the propagation cycle.

\subsection{Data considerations}

In the context of data considerations for a well-functioning agent-based model, it's essential to have detailed data encompassing a wide range of financial aspects. Much of this data can be derived from the multi-layer network:

\begin{itemize}
\item Agents within the framework are characterized by their balance sheets, which are readily available and accessible in the multi-layered network, thanks to the enrichment process outlined in section~\ref{mod4}. This includes information about the risk-weighted assets of the agents, allowing for the computation of their capital ratios.

\item The model's long-term interbank market and short-term funding market can be effectively modeled using adjacency matrices \(W^{l},\ \ l \in G\) within the network. These matrices establish the connections between agents participating in these markets.

\item For modeling fire sales in the securities market, we can start by examining the external securities layer, of which the security-
level representation aligns with the holding matrix \(S \in \mathbb{R}^{n \times m}\) of the model.

\item The price vector \(P\) is implicitly embedded within the multi-layer network, as the external securities layer includes prices for the securities, necessary for calculating the market value of portfolio investments.

\end{itemize}

However, it's worth noting that certain information isn't naturally present in the network:

\begin{itemize}
\item The agent-based model relies on risk weights for both interbank assets and external securities, which are not directly available in the network. A workaround involves using ratings data to attribute risk weights based on the security's rating. Alternatively, constant values for the weights can be assigned.

\item To model price depreciations of securities, it's essential to quantify the liquidity or narrowness of the market for various securities, denoted as \(\alpha \in \mathbb{R}^{m}\). This calibration process typically relies on publicly available market data, including price returns and traded volumes, such as the approach described in \cite{Busseti_2012}. This information is currently not captured by the multi-layer network and needs to be simulated to explore different market scenarios and their potential impact on the interbank system.
\end{itemize}

\end{document}